\documentclass[preprint,12pt]{elsarticle}
\usepackage[T1]{fontenc}
\usepackage[utf8]{inputenc}
\usepackage{CJKutf8}
\usepackage{amsmath,amssymb}
\usepackage{bm}
\usepackage{graphicx}
\usepackage{array}
\usepackage{booktabs}
\usepackage{threeparttable}
\usepackage{float}
\usepackage{enumitem}
\usepackage{tikz}
\usepackage{hyperref}  
\usetikzlibrary{shapes.geometric, arrows, positioning, shadows, calc}
\usepackage{cleveref}
\crefname{equation}{Eq.}{Eqs.}      \Crefname{equation}{Eq.}{Eqs.}
\crefname{figure}{Fig.}{Figs.}      \Crefname{figure}{Fig.}{Figs.}
\crefname{table}{Table}{Tables}     \Crefname{table}{Table}{Tables}
\crefname{section}{Section}{Sections}
\Crefname{section}{Section}{Sections}
\crefformat{appendix}{#2#1#3}       \Crefformat{appendix}{#2#1#3}
\crefformat{subappendix}{#2#1#3}    \Crefformat{subappendix}{#2#1#3}

\journal{Materials Today Physics}

\begin{document}

\begin{frontmatter}

\title{Formulations of  elastodynamic equations for anisotropic multiphase porous
piezoelectric media based on global energy conservation}

\author[a,b]{Xiuming Wang}
\author[a,b]{Yinqiu Zhou \corref{cor1}}
\ead{zhouyinqiu@mail.ioa.ac.cn}
\author[a,b]{Zhixiang Sun}
\author[a,b]{Lin Liu}

\cortext[cor1]{Corresponding author.}

\affiliation[a]{organization={State Key Laboratory of Acoustics, Institute of
                              Acoustics, Chinese Academy of Sciences},
                city={Beijing},
                postcode={100190},
                country={China}}
\affiliation[b]{organization={School of Physical Sciences, University of
                              Chinese Academy of Sciences},
                city={Beijing},
                postcode={100049},
                country={China}}
\begin{abstract}
Multiphase porous piezoelectric media are essential for advanced transducers and smart sensors. Existing theories typically postulate Newton's second law for each phase or rely on phenomenological Hamiltonian constructions. The former forces \emph{ad hoc} virtual-mass tensors to describe interphase inertia, while the latter provides no intrinsic safeguard against thermodynamic inconsistency when piezoelectric and multiphase couplings are superposed. In this work, we establish a linear dynamic and constitutive theory for anisotropic multiphase porous piezoelectric media from global energy conservation (GEC). From an abstract energy density functional, Taylor expansion and symmetry constraints derive the standard kinetic and potential energy densities and electric enthalpy, rather than assuming them a priori. Localization of the GEC integral yields the multiphase momentum equations, Gauss's law, the coupled constitutive relations, and the boundary conditions as mathematical corollaries, without invoking Newton's law or Hamilton's principle. The framework eliminates virtual-mass parameters entirely: interphase inertial coupling emerges organically from the off-diagonal kinetic-energy coefficients $\rho_{ij}^{\alpha\beta}$. Because all coefficients derive from a single smooth potential, Schwarz's theorem automatically guarantees Maxwell reciprocity and full thermodynamic self-consistency. The formulations agree with those from Hamilton's principle and reduce exactly to Biot's poroelastic theory and Tiersten's single-phase piezoelectric theory in the respective limits. Finally, linear plane-wave analysis produces a generalized Christoffel eigenvalue equation, and numerical phase-velocity calculations for water-saturated porous PZT-2 illustrate the modal structures and reveal strongly directional electromechanical coupling.
\end{abstract}
\end{frontmatter}
\section{Introduction}\label{sec:intro}
Multiphase porous piezoelectric media are complex media composed of a solid skeleton and pore fluids (for example, liquid and gas phases), in which the solid skeleton exhibits piezoelectricity \cite{Rohan2018, Tariq2025, Kudimova2017, Arai1991, Kar2007}. Examples include piezoceramic--polymer composites, fluid-saturated piezoelectric bone tissue, and porous piezoelectric metamaterials. An accurate description of the coupled elastic-wave propagation in these media, which involves the mechanical, fluid, and electric fields, is therefore essential for nondestructive material characterization, biomedical ultrasound diagnosis, underwater acoustic transducer design, and smart sensor development \cite{Rohan2018, Bowen2022, Tezcan2025}. Since Biot (1956) established the theory of elastic waves in fluid-saturated porous media \cite{Biot1956}, multiphase continuum mechanics has advanced considerably. Coussy (2004) systematized the theory and extended it to complex poromechanics \cite{Coussy2004}, and further extensions and applications can be found in refs. \cite{Santos1990,Leclaire1994, Leclaire1995}. Meanwhile, the continuum theory of piezoelectric media has also matured: Tiersten (1969) established a systematic constitutive and wave-propagation framework for linear piezoelectric continua \cite{Tiersten1969}, building on the broader foundations of continuum field theory and of linear elasticity provided by Truesdell and Toupin (1960) and by Gurtin (1972) \cite{Truesdell1960, Gurtin1972}. However, when one attempts to combine porous-media theory with piezoelectricity and to extend it to the coexistence of multiple fluid phases, the traditional theoretical approaches face considerable challenges.

First, the limitations of postulating Newton's second law. Most existing studies postulate Newton's second law and describe multiphase dynamics by writing a separate momentum balance for each phase. In multiphase media, however, a direct application of Newton's second law requires the introduction of a ``virtual mass'' (added mass) tensor to represent the interphase inertial coupling \cite{Biot1956b, Bedford1984}. In essence, the virtual-mass effect arises because an accelerating dispersed phase (bubbles, droplets, or solid particles) must simultaneously drag the surrounding continuous phase along with it; this effect is difficult to model naturally within the framework of continuum mechanics \cite{Capriz1987}. These coefficients depend on the pore geometry and require constitutive or microstructural determination \cite{Bedford1984, Yavari1988}. Consequently, the traditional route that presupposes Newton's second law must supplement it with additional interphase inertial parameters that the balance law itself does not determine.

Second, the reliance on variational principles and its limitations. Mainstream continuum theories rely heavily on Hamilton's principle \cite{Tiersten1969, Cassel2013}. Although the variational route is mathematically elegant, constructing a globally valid Lagrangian functional is often phenomenological and nontrivial for systems involving complex multiphase interfaces, non-conservative forces, or specific dissipation mechanisms. The variational principle is essentially another mathematical statement of energy conservation \cite{Zhou2022a, Wang2023, Zhou2022b}, but its applicability is often limited when nonstandard multiphase coupling boundary conditions are involved.

Third, the phenomenological character of the constitutive relations and the risk of thermodynamic inconsistency. In the theoretical construction of multiphase porous piezoelectric media, superposing multiphase coupling and piezoelectricity in an isolated manner constitutes a typical ``bottom-up'' patchwork strategy. This strategy faces a double difficulty. Theoretically, it lacks a unified thermodynamic potential as a constraint, so the resulting coupled constitutive relations are not guaranteed to satisfy thermodynamic restrictions such as the Maxwell reciprocity relations; by contrast, thermodynamically constrained theories of porous piezoelectric materials impose such restrictions explicitly and derive the corresponding reciprocity relations \cite{Ciarletta1993, Ciarletta1996}. Practically, specific conventional approximations have been shown to violate the second law of thermodynamics \cite{Wilhelmsen2024}, and free-energy-based formulations have been introduced to restore thermodynamic consistency and energy stability \cite{Kou2026}. Recent work also illustrates a top-down thermodynamic construction for broadly coupled geological processes \cite{Yarushina2026}.

To overcome these limitations, an alternative route of theoretical construction has emerged in recent years: taking the principle of energy conservation as the starting point, rather than postulating Newton's law or relying on variational principles, to derive the dynamic equations and constitutive relations of continuum mechanics. The conceptual basis of this idea can be traced back to the general discussion of energy methods in continuum field theory by Truesdell and Toupin (1960) \cite{Truesdell1960}.
Recently, we have systematically investigated how wave-dynamic equations can be established within an energy-conservation framework \cite{Zhou2022a,Zhou2022b,Wang2023}. These studies show how the energy-conservation method proceeds from a global energy balance and, through mathematical localization, derives the dynamic equations of discrete media and of isotropic elastic media. They also examine the key mathematical steps required when the method is extended to anisotropic media and to multiphase porous media. Most recently, we further extended the energy-conservation framework to multiscale wave-induced fluid flow in partially saturated porous media \cite{Liu2026}, and analysed multi-field coupling in piezoelectric elastic media \cite{Zhou2021}. These works laid the methodological foundation for the present study. They focused, however, on purely elastic inhomogeneous anisotropic media and on multiphase porous media, and did not consider media containing both a piezoelectric skeleton and multiple coexisting fluid phases.

The above works indicate that the principle of energy conservation is not merely a tool for checking theoretical consistency; it can serve as an independent logical starting point for constructing continuum dynamic theories. The methodology has been developed across several settings: from isotropic to anisotropic media, from single-phase elasticity to multiphase porous media, and from conservative to dissipative systems.

The dynamic equations of porous piezoelectric media have been explored along several theoretical routes. Within a linear thermodynamic framework, Ciarletta and Scalia (1993) derived the basic equations of the linear theory of porous piezoelectric materials using the entropy-production inequality, and discussed the thermodynamic restrictions on the constitutive equations \cite{Ciarletta1993}. Ciarletta and Scarpetta (1996) further adopted a generalized thermodynamic method to derive the constitutive equations and thermodynamic restrictions of thermoelastic porous piezoelectric materials, and established uniqueness theorems and variational principles \cite{Ciarletta1996}. At the micromechanical level, Rohan and Luke\v{s} (2018) applied asymptotic homogenization to derive the macroscopic constitutive behaviour of fluid-saturated piezoelectric porous media from periodic microstructures, revealing the influence of pore-scale fluid--solid coupling on the macroscopic effective properties \cite{Rohan2018}. Recently, we proposed a methodology for constructing the dynamic equations of thermo-piezoelectric dissipative media from the principle of energy conservation; the resulting equations are consistent with those obtained from Hamilton's principle \cite{Zhou2021}. These works, from different perspectives, laid the foundation for the theoretical modelling of multiphase porous piezoelectric media and motivate the energy-conservation construction developed in this paper.

In summary, existing studies have not yet systematically treated the electromechanical coupling problem of anisotropic media in which piezoelectricity and multiple fluid phases are present simultaneously. When piezoelectricity is introduced into a multiphase porous medium, the energy density functional of the system contains not only mechanical variables (displacement, strain, velocity) but also electrical variables (the electric potential and the electric field). The cross-coupling terms among these variables are constrained by time-reversal symmetry and spatial inversion symmetry: the former forbids cross terms containing an odd number of velocities, while the latter determines under which crystal symmetries the electromechanical coupling terms can be nonzero. These two symmetry constraints make the selection of admissible terms in the energy density functional, and the derivation of the constitutive relations, considerably more involved than in the purely elastic or poroelastic case. Moreover, when more than two fluid phases coexist in the system, the structure of the constitutive coupling coefficients among the fluid phases, and between each fluid phase and the piezoelectric solid skeleton, becomes richer, and their influence on the number of propagating wave modes and on the dispersion characteristics becomes more complex.

Motivated by these considerations, this paper takes the first law of thermodynamics (or global energy conservation) as its starting point and constructs a linear dynamic and constitutive theory of anisotropic and multiphase porous piezoelectric media. Starting from the general formula of the multivariate Taylor expansion, we first derive the energy density functional of the multiphase system order by order, and use the natural-equilibrium condition and time-reversal symmetry to eliminate the inadmissible terms (for example, cross terms containing an odd number of velocities). Then, through mathematical localization of the energy conservation integral, the multiphase momentum equations, Gauss's law, and the coupled constitutive relations emerge naturally as mathematical corollaries, which guarantees the thermodynamic self-consistency of the theory at its root. Finally, we investigate linear plane-wave propagation in such media and derive a generalized Christoffel eigenvalue equation with multiphase coupling, providing a theoretical basis for the acoustic characterization of multiphase porous piezoelectric media.

Notation. Throughout this paper, the Einstein summation convention is adopted: every repeated index, including the spatial indices $i,j,k,l,m,n = 1,2,3$ and the phase indices $\alpha,\beta = 1,2,\dots, M$, implies summation over all of its admissible values, and no explicit summation symbol is written.

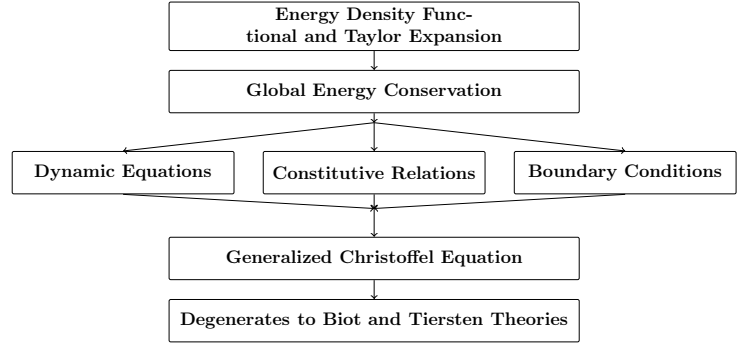
\begin{figure}[H]
\centering
\resizebox{0.70\textwidth}{!}{%
\begin{tikzpicture}[
    node distance=0.4cm and 0.6cm,
    box/.style={rectangle, draw=black, line width=0.6pt, fill=white, text=black,
                text width=4.5cm, align=center, rounded corners=1pt,
                minimum height=0.9cm, inner sep=3pt,
                font=\small\bfseries},
    widebox/.style={rectangle, draw=black, line width=0.6pt, fill=white, text=black,
                    text width=8.5cm, align=center, rounded corners=1pt,
                    minimum height=0.9cm, inner sep=3pt,
                    font=\small\bfseries}
]
\node[widebox] (A) {Energy Density Functional and Taylor Expansion};
\node[widebox, below=of A] (B) {Global Energy Conservation};
\node[box, below=0.8cm of B] (C2) {Constitutive Relations};
\node[box, left=of C2] (C1) {Dynamic Equations};
\node[box, right=of C2] (C3) {Boundary Conditions};
\coordinate (split) at ($(B.south) + (0,-0.2cm)$);
\coordinate (merge) at ($(C1.south)!0.5!(C3.south) + (0,-0.3cm)$);
\node[widebox, below=0.6cm of merge] (D) {Generalized Christoffel Equation};
\node[widebox, below=of D] (E) {Degenerates to Biot and Tiersten Theories};
\draw[->, line width=0.6pt] (A) -- (B);
\draw[->, line width=0.6pt] (B.south) -- (split);
\draw[->, line width=0.6pt] (split) -- (C1.north);
\draw[->, line width=0.6pt] (split) -- (C2.north);
\draw[->, line width=0.6pt] (split) -- (C3.north);
\draw[->, line width=0.6pt] (C1.south) -- (merge);
\draw[->, line width=0.6pt] (C2.south) -- (merge);
\draw[->, line width=0.6pt] (C3.south) -- (merge);
\draw[->, line width=0.6pt] (merge) -- (D.north);
\draw[->, line width=0.6pt] (D) -- (E);
\end{tikzpicture}}
\caption{The development roadmap for the theoretical framework.}
\label{fig:roadmap}
\end{figure}

\Cref{fig:roadmap} summarises the logical structure of the theory developed in this paper. Starting from the energy density functional and its Taylor expansion, the global energy balance is localised to yield, in parallel, the dynamic equations, the constitutive relations and the boundary conditions; these in turn lead to the generalized Christoffel equation, which degenerates to the Biot and Tiersten theories in the corresponding limits.

\section{Axiomatic basis}\label{sec:axioms}

A continuum theory should be built on an explicit system of axioms and basic assumptions. To make the derivations of this paper traceable, this section lists the axiomatic premises and basic assumptions on which the theory relies. These premises form the logical basis of the subsequent Taylor expansion, symmetry-based term selection, and localization of energy conservation.

\subsection{Energy conservation law}\label{sec:energy_law}

The energy conservation law (the first law of thermodynamics) is a universal principle of continuum mechanics. For any kinematically admissible process (satisfying continuity, compatibility, and so on), the rate of change of the total energy of the system (the sum of the kinetic, strain, electric, and other forms of energy) equals the sum of the power of the external forces and the electromagnetic power \cite{Truesdell1960, Gurtin1981, Malvern1969}.

Here ``energy'' is understood as a function of the state of the system: its value is completely determined by the state variables (such as displacement, velocity, strain, and electric field) and is independent of the historical path by which the system reached that state \cite{Callen1985}. In other words, for a given set of state variables there exists a unique energy density functional, whose form is determined by the material state and by the prescribed external sources.

The mathematical statement of this principle can be given at three levels, according to the way in which the system exchanges energy with its surroundings.

\medskip
\noindent\textbf{(a) Isolated system}

For an isolated system that exchanges no energy with its surroundings, the total energy $\mathcal E_{\rm tot}$ of the system remains constant \cite{Landau1976, Callen1985}:
\begin{equation}
\frac{{\rm d}\mathcal E_{\rm tot}}{{\rm d}t} = 0, \qquad \mathcal E_{\rm tot} = {\rm const}.
\label{eq:closed}
\end{equation}

\medskip
\noindent\textbf{(b) System with external energy exchange (general thermodynamic form)}

For a system that exchanges energy with its surroundings, the rate of change of the total energy equals the sum of the rate of work $\dot W$ done on the system by the surroundings and the heat-transfer rate $\dot Q$ \cite{Callen1985, Cengel2002}:
\begin{equation}
\frac{{\rm d}\mathcal E_{\rm tot}}{{\rm d}t} = \dot W + \dot Q.
\label{eq:open}
\end{equation}
The piezoelectric continuum considered in this paper belongs to this class of systems; its energy exchange consists of mechanical work (done by body and surface forces) and electrical work (done by charges).

\medskip
\noindent\textbf{(c) Global integral form for multiphase piezoelectric continua}

Specifically, for an arbitrary spatial volume $V$ with boundary $\partial V$, the global integral form of the energy conservation law for the present problem reads
\begin{equation}
\frac{{\rm d}}{{\rm d}t} \int_V \mathcal{E} \, {\rm d}V = \int_V f_i^\alpha \dot u_i^\alpha {\rm d}V + \int_{\partial V} \left( t_i^\alpha \dot u_i^\alpha + \sigma_e \dot\varphi \right) {\rm d}S,
\label{eq:energy_axiom}
\end{equation}
where $\mathcal{E}$ denotes the augmented energy-density functional in the electric-field representation, including the prescribed-source coupling $\rho_f\varphi$; $f_i^\alpha$ and $t_i^\alpha$ are the body force and the surface traction acting on phase $\alpha$, and $\sigma_e$ is the surface charge density, respectively. This equation is the concrete form taken by the first law of thermodynamics in a multiphase piezoelectric continuum. Its general axiomatic form can be found in continuum mechanics \cite{Truesdell1960, Gurtin1981, Malvern1969}, its form applicable to porous media in ref. \cite{Coussy2004}, and its use as a first principle for deriving dynamic equations in ref. \cite{Zhou2021}.

The physical meaning of \cref{eq:energy_axiom} can be read term by term as follows:
\begin{itemize}[leftmargin=*]
\item \textbf{Left-hand side}: the rate of change of the augmented energy-density functional $\mathcal{E}$, which includes the prescribed-source coupling $\rho_f\varphi$, contained in the volume $V$ \cite{Truesdell1960};
\item \textbf{First term on the right}: the mechanical power supplied by the body forces acting through the velocity of each phase (gravity, electromagnetic body forces, and so on) \cite{Gurtin1981};
\item \textbf{Second term on the right}: the mechanical power supplied by the external surface tractions of each phase, plus the electrical power supplied by the free surface charge on the boundary through the rate of change of the electric potential \cite{Jackson1999}.
\end{itemize}

As the axiom of the present theory, this equation presupposes neither Newton's second law nor any variational principle. The dynamic equations (the momentum equations and Gauss's law) and the constitutive relations can be derived from its mathematical localization. In deriving the field equations and the natural boundary conditions, this energy-conservation route shares the same mathematical structure as the variational route \cite{Truesdell1960, Tiersten1969, Gurtin1981}.

\subsection{Mathematical assumptions}\label{sec:assumptions}

The following mathematical assumptions are adopted based on the above axiom.
\begin{enumerate}[label=(\alph*), leftmargin=*, align=left]
  \item \textbf{Energy density}: the system possesses an energy density functional consisting of a material part and the coupling to the prescribed free-charge source,
  \[
  \mathcal{E} = \mathcal{E}_{\rm m}(\dot u_i^\alpha, \varepsilon_{ij}^\alpha, E_i, x_i) + \rho_f \varphi ,
  \]
  where $\mathcal{E}_{\rm m}$ is of class $C^\infty$ in a neighbourhood of the origin. The free volume charge density $\rho_f$ is treated as a prescribed, time-independent source, so that $\dot\rho_f = 0$ and $\delta\rho_f = 0$.

  \item \textbf{Objectivity}: the elastic--electric storage part of $\mathcal{E}_{\rm m}$ is objective, so its dependence on the displacement gradients enters only through $\varepsilon_{ij}^\alpha$; the velocities entering the kinetic part are measured in a fixed inertial frame:
  \[
  \mathcal{E}_{\rm m} = \mathcal{E}_{\rm m}(\dot u_i^\alpha, \varepsilon_{ij}^\alpha, E_i).
  \]

  \item \textbf{Time-reversal invariance}: $\mathcal{E}_{\rm m}(\dot u, \varepsilon, E) = \mathcal{E}_{\rm m}(-\dot u, \varepsilon, E)$, since $\varepsilon$, $E$, and $x$ are even under time reversal.

  \item \textbf{Natural equilibrium state}: at $(\dot u_i^\alpha, \varepsilon_{ij}^\alpha, E_i) = (0,0,0)$ the first variation of the material part vanishes:
  \[
  \left.\frac{\partial\mathcal{E}_{\rm m}}{\partial \dot u_i^\alpha}\right|_0 = 0,\quad
  \left.\frac{\partial\mathcal{E}_{\rm m}}{\partial \varepsilon_{ij}^\alpha}\right|_0 = 0,\quad
  \left.\frac{\partial\mathcal{E}_{\rm m}}{\partial E_i}\right|_0 = 0.
  \]
  The electric field is quasi-static: $E_i = -\varphi_{,i}$, i.e., radiation effects are neglected.

  \item \textbf{Continuum hypothesis}: the wavelength $\lambda$ is much larger than the microstructural scale, and $\|u_{i,j}^\alpha\|\ll 1$; the strain is defined as $\varepsilon_{ij}^\alpha = (u_{i,j}^\alpha + u_{j,i}^\alpha)/2$, and the problem is linear.
\end{enumerate}

\subsection{Second-order Taylor expansion}\label{sec:taylor}

By the energy-density assumption of the axiomatic basis ($C^\infty$ smoothness), the energy density is now extended to depend explicitly on the electric potential $\varphi$ and on the spatial coordinates $x_i$, in addition to the velocity, strain, and electric field:
\begin{equation}
\mathcal{E} = \mathcal{E}_{\rm m}(\dot{u}_i^\alpha,\; \varepsilon_{ij}^\alpha,\; E_i) + \rho_f \varphi,
\label{eq:E_dep}
\end{equation}
where $E_i = -\varphi_{,i}$. In the neighbourhood of the natural equilibrium state $(\dot{u}_i^\alpha,\varepsilon_{ij}^\alpha,E_i)=(0,0,0)$, we expand the material part $\mathcal{E}_{\rm m}$ using the multivariate Taylor formula with remainder up to second order. The second-order cross terms containing an odd number of velocities are not displayed here, since they are shown below to vanish by time-reversal symmetry (\cref{eq:tr_forbidden}):
\begin{align}
\mathcal{E}_{\rm m} = \;& \underbrace{\mathcal{E}_{\rm m}(0,0,0)}_{\text{zeroth-order constant}}
+ \underbrace{\begin{aligned}[t]
&\left.\frac{\partial \mathcal{E}_{\rm m}}{\partial \dot{u}_i^\alpha}\right|_0 \dot{u}_i^\alpha
+ \left.\frac{\partial \mathcal{E}_{\rm m}}{\partial \varepsilon_{ij}^\alpha}\right|_0 \varepsilon_{ij}^\alpha \\
&+ \left.\frac{\partial \mathcal{E}_{\rm m}}{\partial E_i}\right|_0 E_i
\end{aligned}}_{\text{first-order terms}} \nonumber \\
& + \frac12
\underbrace{
\left.\frac{\partial^2 \mathcal{E}_{\rm m}}{\partial \dot{u}_i^\alpha \partial \dot{u}_j^\beta}\right|_0 \dot{u}_i^\alpha \dot{u}_j^\beta
}_{\text{kinetic energy}}
+ \frac12
\underbrace{
\left.\frac{\partial^2 \mathcal{E}_{\rm m}}{\partial \varepsilon_{ij}^\alpha \partial \varepsilon_{kl}^\beta}\right|_0 \varepsilon_{ij}^\alpha \varepsilon_{kl}^\beta
}_{\text{elastic energy}} \nonumber \\
& + \frac12
\underbrace{
\left.\frac{\partial^2 \mathcal{E}_{\rm m}}{\partial E_i \partial E_j}\right|_0 E_i E_j
}_{\text{dielectric energy}}
+
\underbrace{
\left.\frac{\partial^2 \mathcal{E}_{\rm m}}{\partial E_i \partial \varepsilon_{jk}^\alpha}\right|_0 E_i \varepsilon_{jk}^\alpha
}_{\text{piezoelectric coupling}}
+ \mathcal{O}(3).
\label{eq:taylor_full}
\end{align}
The natural equilibrium condition eliminates all first-order terms \cite{Malvern1969}. The time-reversal symmetry forbids cross terms containing an odd number of velocities. The remaining second-order coefficients are identified as the material constants:
\begin{align}
& \left.\frac{\partial^2 \mathcal{E}_{\rm m}}{\partial \dot{u}_i^\alpha \partial \dot{u}_j^\beta}\right|_0: = \rho_{ij}^{\alpha\beta}
\quad \text{(multiphase coupled mass density tensor)}, \label{eq:rho_def} \\
& \left.\frac{\partial^2 \mathcal{E}_{\rm m}}{\partial \varepsilon_{ij}^\alpha \partial \varepsilon_{kl}^\beta}\right|_0: = C_{ijkl}^{\alpha\beta}
\quad \text{(multiphase coupled elastic tensor)}, \label{eq:C_def} \\
& \left.\frac{\partial^2 \mathcal{E}_{\rm m}}{\partial E_i \partial \varepsilon_{jk}^\alpha}\right|_0: = -e_{ijk}^\alpha
\quad \text{(multiphase piezoelectric stress tensor)}, \label{eq:e_def} \\
& \left.\frac{\partial^2 \mathcal{E}_{\rm m}}{\partial E_i \partial E_j}\right|_0: = -\kappa_{ij}
\quad \text{(dielectric tensor)}. \label{eq:kappa_def}
\end{align}

Neglecting the higher-order terms and the zeroth-order constant, and restoring the prescribed-source coupling $\rho_f\varphi$ of \cref{eq:E_dep}, the linear energy density of the multiphase porous piezoelectric medium becomes
\begin{equation}
\mathcal{E} = \frac{1}{2} \rho_{ij}^{\alpha\beta} \dot{u}_i^\alpha \dot{u}_j^\beta
+ \frac{1}{2} C_{ijkl}^{\alpha\beta} \varepsilon_{ij}^\alpha \varepsilon_{kl}^\beta
- e_{ijk}^\alpha E_i \varepsilon_{jk}^\alpha
- \frac{1}{2} \kappa_{ij} E_i E_j
+ \rho_f \varphi
\label{eq:energy_linear}
\end{equation}
which is consistent with the standard energy density of linear piezoelectric continua \cite{Tiersten1969} supplemented by the electrostatic interaction term. The corresponding electric enthalpy density (potential energy density) is then defined as
\begin{equation}
\Psi = \mathcal{E} - \frac{1}{2} \rho_{ij}^{\alpha\beta} \dot{u}_i^\alpha \dot{u}_j^\beta - \rho_f \varphi,
\label{eq:psi_def_1}
\end{equation}
thus
\begin{equation}
\Psi = \frac{1}{2} C_{ijkl}^{\alpha\beta} \varepsilon_{ij}^\alpha \varepsilon_{kl}^\beta
- e_{ijk}^\alpha E_i \varepsilon_{jk}^\alpha
- \frac{1}{2} \kappa_{ij} E_i E_j.
\label{eq:psi_def}
\end{equation}

The inclusion of the $\rho_f \varphi$ term ensures that the total energy accounts for the work done by the external charge distribution, and it will be crucial for deriving the standard form of Gauss's law in the subsequent localization procedure.

The linear energy density obtained in \cref{eq:energy_linear} contains only five types of terms: kinetic energy, elastic energy, piezoelectric coupling, dielectric energy, and the charge--potential interaction energy $\rho_f\varphi$. All other terms that appear in the full Taylor expansion are excluded by physical principles or by the linearization assumption. We now explain the reasoning term by term.

The selection rules are as follows.

\medskip
\noindent\textbf{First-order terms.}
At the natural equilibrium state, the system is free of initial stress, remanent polarization, and motion, so the first variations with respect to the dynamic variables vanish \cite{Malvern1969}:
\begin{equation}
\left.\frac{\partial \mathcal{E}_{\rm m}}{\partial \dot u_i^\alpha}\right|_0 = 0,\qquad
\left.\frac{\partial \mathcal{E}_{\rm m}}{\partial \varepsilon_{ij}^\alpha}\right|_0 = 0,\qquad
\left.\frac{\partial \mathcal{E}_{\rm m}}{\partial E_i}\right|_0 = 0.
\label{eq:first_order_vanish}
\end{equation}
Thus all first-order terms of the material energy are discarded.

\medskip
\noindent\textbf{Second-order self-coupling terms.}
The three quadratic terms $\dot u_i^\alpha \dot u_j^\beta$, $\varepsilon_{ij}^\alpha \varepsilon_{kl}^\beta$, and $E_i E_j$ are the fundamental energy contributions of any linear piezoelectric system: kinetic energy, elastic strain energy, and dielectric field energy, respectively \cite{Tiersten1969, Landau1960}. They are all retained and their coefficients define the mass density tensor $\rho_{ij}^{\alpha\beta}$, the elastic tensor $C_{ijkl}^{\alpha\beta}$, and the dielectric tensor $\kappa_{ij}$.

\medskip
\noindent\textbf{Second-order cross terms and time-reversal symmetry.}
The system is conservative and satisfies time-reversal symmetry ($t\to -t$). Under this transformation, the velocity changes sign ($\dot u_i^\alpha \to -\dot u_i^\alpha$), while the strain and the electric field are even ($\varepsilon_{ij}^\alpha \to +\varepsilon_{ij}^\alpha$, $E_i \to +E_i$). Since the material energy density $\mathcal{E}_{\rm m}$ is a scalar invariant under time reversal, it must be an even function of the velocities. Consequently, any cross term containing an odd number of velocities is forbidden \cite{Landau1960, Zhou2022a}:
\begin{equation}
\left.\frac{\partial^2 \mathcal{E}_{\rm m}}{\partial \dot u_i^\alpha \partial \varepsilon_{jk}^\beta}\right|_0 = 0,\qquad
\left.\frac{\partial^2 \mathcal{E}_{\rm m}}{\partial \dot u_i^\alpha \partial E_j}\right|_0 = 0.
\label{eq:tr_forbidden}
\end{equation}
The remaining cross term $E_i \varepsilon_{jk}^\alpha$ is even under time reversal and is therefore allowed; it represents the piezoelectric coupling and is retained, with its coefficient identified as $-e_{ijk}^\alpha$ \cite{Tiersten1969}.

\medskip
\noindent\textbf{The prescribed-source coupling.}
The prescribed-source coupling $\rho_f\varphi$ is added separately and is not part of the Taylor expansion of the material energy \cite{Tiersten1969, Maugin1988}.

\medskip
Applying these selection rules, the surviving terms of the material expansion are the first four terms of \cref{eq:energy_linear}; adding the prescribed-source coupling $\rho_f\varphi$ gives the complete augmented energy-density functional. The material part is consistent with the standard energy density of linear piezoelectric continua \cite{Tiersten1969}, and the source term accounts for the work done by the prescribed external charges \cite{Landau1960, Jackson1999}.

Although the present work is restricted to linear conservative media, the framework is inherently extendable to nonlinear regimes by retaining higher-order terms in the energy expansion, and to viscoelastic behaviour by relaxing the time-reversal symmetry condition, which allows velocity--strain and velocity--field coupling terms in the energy density to naturally produce rate-dependent constitutive relations upon localization of the energy balance. This provides a unified first-principles pathway for modelling nonlinear dissipative multiphase piezoporous media.

\subsection{Localization based on global energy conservation}\label{sec:localization}

Because the interaction energy $\rho_f \varphi$ is already contained in the left-hand side of the energy balance, the global energy conservation equation (the first law of thermodynamics) now takes the form
\begin{equation}
\frac{{\rm d}}{{\rm d}t} \int_V \mathcal{E} \, {\rm d}V
=\int_V f_i^\alpha \dot{u}_i^\alpha \, {\rm d}V
+\int_{\partial V} (t_i^\alpha \dot{u}_i^\alpha + \sigma_e \dot{\varphi}) \, {\rm d}S
\label{eq:global_energy}
\end{equation}
where the electrical power from free volume charges no longer appears explicitly on the right-hand side; it is implicitly included through the time derivative of $\rho_f \varphi$ in $\mathcal{E}$. This is the correct form of the energy balance for a system whose energy functional contains the electrostatic potential energy \cite{Landau1960, Jackson1999}.

Substituting $\mathcal{E}$ from \cref{eq:energy_linear} into the left-hand side of \cref{eq:global_energy}, and using $E_i = -\varphi_{,i}$, we obtain
\begin{equation}
\begin{aligned}
\frac{{\rm d}}{{\rm d}t} \int_V \mathcal{E} \, {\rm d}V
= \;& \int_V \rho_{ij}^{\alpha\beta} \ddot{u}_j^\beta \dot{u}_i^\alpha \, {\rm d}V
+ \int_V \frac{\partial \Psi}{\partial \varepsilon_{ij}^\alpha} \dot{u}_{i,j}^\alpha \, {\rm d}V \\
& - \int_V \frac{\partial \Psi}{\partial E_i} \dot{\varphi}_{,i} \, {\rm d}V
+ \int_V \rho_f \dot{\varphi} \, {\rm d}V .
\end{aligned}
\label{eq:energy_rate}
\end{equation}

The term $\dot u_{i,j}^\alpha$ appearing in the rate of change of the potential energy is a direct consequence of the chain rule and the linear strain--displacement relation. Under the small-deformation assumption, the strain rate is
\begin{equation}
\dot{\varepsilon}_{ij}^\alpha = \frac12 (\dot u_{i,j}^\alpha + \dot u_{j,i}^\alpha),
\label{eq:strain_rate}
\end{equation}
where $\dot u_{i,j}^\alpha$ is the gradient of the velocity field. Since $\partial\Psi/\partial\varepsilon_{ij}^\alpha$ is symmetric ($\partial\Psi/\partial\varepsilon_{ij}^\alpha =\partial\Psi/\partial\varepsilon_{ji}^\alpha$), the contraction with the strain rate reduces to \cite{Malvern1969}
\begin{equation}
\frac{\partial \Psi}{\partial \varepsilon_{ij}^\alpha} \dot \varepsilon_{ij}
= \frac{\partial \Psi}{\partial \varepsilon_{ij}^\alpha} \dot u_{i,j}^\alpha .
\label{eq:sym_contraction}
\end{equation}
This replacement is an exact consequence of the symmetry of the stress tensor and the definition of the infinitesimal strain. The resulting term $\partial\Psi/\partial\varepsilon_{ij}^\alpha \,\dot u_{i,j}^\alpha$ is then ready for integration by parts, which converts it into a boundary traction term and a volume divergence term that ultimately yields the momentum balance equation.

Moving all terms to the left-hand side gives
\begin{equation}
\begin{aligned}
& \int_V \rho_{ij}^{\alpha\beta} \ddot{u}_j^\beta \dot{u}_i^\alpha \, {\rm d}V
+ \int_V \frac{\partial \Psi}{\partial \varepsilon_{ij}^\alpha} \dot{u}_{i,j}^\alpha \, {\rm d}V
- \int_V \frac{\partial \Psi}{\partial E_i} \dot{\varphi}_{,i} \, {\rm d}V \\
& + \int_V \rho_f \dot{\varphi} \, {\rm d}V
- \int_V f_i^\alpha \dot{u}_i^\alpha \, {\rm d}V
- \int_{\partial V} t_i^\alpha \dot{u}_i^\alpha \, {\rm d}S
- \int_{\partial V} \sigma_e \dot{\varphi} \, {\rm d}S = 0 .
\end{aligned}
\label{eq:moved}
\end{equation}
Applying the divergence theorem to the terms involving spatial derivatives:
\begin{align}
\int_V \frac{\partial \Psi}{\partial \varepsilon_{ij}^\alpha} \dot{u}_{i,j}^\alpha \, {\rm d}V
&= \int_{\partial V} \frac{\partial \Psi}{\partial \varepsilon_{ij}^\alpha} n_j \dot{u}_i^\alpha \, {\rm d}S \nonumber\\
&\quad - \int_V \left( \frac{\partial \Psi}{\partial \varepsilon_{ij}^\alpha} \right)_{\!,j} \dot{u}_i^\alpha \, {\rm d}V, \label{eq:div_stress} \\
- \int_V \frac{\partial \Psi}{\partial E_i} \dot{\varphi}_{,i} \, {\rm d}V
&= - \int_{\partial V} \frac{\partial \Psi}{\partial E_i} n_i \dot{\varphi} \, {\rm d}S \nonumber\\
&\quad + \int_V \left( \frac{\partial \Psi}{\partial E_i} \right)_{\!,i} \dot{\varphi} \, {\rm d}V, \label{eq:div_D}
\end{align}
and, inserting \cref{eq:div_stress,eq:div_D} into \cref{eq:moved} and collecting volume and surface integrals separately yields the energy balance:
\begin{equation}
\begin{split}
& \int_V \left[
\rho_{ij}^{\alpha\beta} \ddot{u}_j^\beta
- \left( \frac{\partial \Psi}{\partial \varepsilon_{ij}^\alpha} \right)_{\!,j}
- f_i^\alpha
\right] \dot{u}_i^\alpha \, {\rm d}V \\
& + \int_V \left[
\left( \frac{\partial \Psi}{\partial E_i} \right)_{\!,i}
+ \rho_f
\right] \dot{\varphi} \, {\rm d}V
+ \int_{\partial V} \left(
\frac{\partial \Psi}{\partial \varepsilon_{ij}^\alpha} n_j
- t_i^\alpha
\right) \dot{u}_i^\alpha \, {\rm d}S \\
& - \int_{\partial V} \left(
\frac{\partial \Psi}{\partial E_i} n_i
+ \sigma_e \right) \dot{\varphi} \, {\rm d}S = 0.
\end{split}
\label{eq:localized}
\end{equation}

Since the velocity fields $\dot{u}_i^\alpha$ and the potential rate $\dot{\varphi}$ are arbitrary and independent in the volume and on the boundary, the fundamental lemma of the calculus of variations \cite{Gelfand1963, Lanczos1970} requires that their coefficients vanish separately. This leads to the volume equations:
\begin{equation}
\rho_{ij}^{\alpha\beta} \ddot{u}_j^\beta= \left( \frac{\partial \Psi}{\partial \varepsilon_{ij}^\alpha} \right)_{\!,j}
+ f_i^\alpha,
\label{eq:pre_momentum}
\end{equation}
\begin{equation}
 \left( \frac{\partial \Psi}{\partial E_i} \right)_{\!,i}+\rho_f = 0,
\label{eq:pre_gauss}
\end{equation}
and the boundary conditions:
\begin{align}
\frac{\partial \Psi}{\partial \varepsilon_{ij}^\alpha} n_j &= t_i^\alpha, \label{eq:pre_traction} \\
- \frac{\partial \Psi}{\partial E_i} n_i &= \sigma_e. \label{eq:pre_charge}
\end{align}

\Cref{eq:localized} is the central result of the localization procedure. It contains, in a single integral identity, all the volume and boundary information of the system. Importantly, this equation has been obtained entirely from the global energy balance without invoking Hamilton's variational principle or postulating Newton's second law. The subsequent field equations and natural boundary conditions emerge merely as a consequence of the arbitrariness of the test functions $\dot{u}_i^\alpha$ and $\dot{\varphi}$ in the volume and on the boundary. Thus, \cref{eq:localized} serves as the unified ``mother equation'' from which both the governing equations and the boundary conditions follow as mathematical corollaries. This demonstrates that global energy conservation alone is sufficient to generate the complete set of continuum field equations, including the multiphase momentum balances, Gauss's law, and the associated traction and charge boundary conditions. This procedure is consistent with those in refs. \cite{Zhou2021,Zhou2022b}.

\subsubsection{Elastodynamic equations, Gauss's law, and boundary conditions}\label{sec:field_eqs}

Now we introduce the standard definitions of the Cauchy stress tensor and the electric displacement as thermodynamic conjugates \cite{Tiersten1969, Maugin1988}:
\begin{equation}
\sigma_{ij}^\alpha: = \frac{\partial \Psi}{\partial \varepsilon_{ij}^\alpha}, \qquad
D_i: = -\frac{\partial \Psi}{\partial E_i},
\label{eq:stress_D_def}
\end{equation}
and these definitions are purely a naming convention; they arise from the boundary terms of the energy conservation equation. The stress tensor $\sigma_{ij}^\alpha$ represents the mechanical force per unit area acting on the $\alpha$-th phase, while the electric displacement $D_i$ represents the electric flux density. The minus sign in the definition of $D_i$ originates from the Legendre transformation that changes the independent variable from the electric displacement to the electric field \cite{Landau1960}.

Substituting these definitions into \cref{eq:pre_momentum,eq:pre_gauss,eq:pre_traction,eq:pre_charge} gives the final governing equations and boundary conditions.

\medskip
From \cref{eq:pre_momentum}, we obtain the multiphase momentum balance equation:
\begin{equation}
\rho_{ij}^{\alpha\beta} \ddot{u}_j^\beta - \sigma_{ij,j}^\alpha - f_i^\alpha = 0.
\label{eq:momentum}
\end{equation}
This equation is the continuum analogue of Newton's second law for each phase: the inertial force (left-hand side) balances the divergence of the stress tensor (internal forces) plus the external body force. Notably, it has been derived from energy conservation rather than being postulated a priori. The off-diagonal blocks $\rho_{ij}^{\alpha\beta}$ with $\alpha \neq \beta$ naturally describe the interphase inertial coupling, replacing the phenomenological ``virtual mass'' parameters of conventional theories.

\medskip
From \cref{eq:pre_gauss}, we obtain Gauss's law:
\begin{equation}
D_{i,i} = \rho_f.
\label{eq:gauss}
\end{equation}
This is the standard Maxwell equation for the electric displacement in the presence of free volume charges. Its positive sign has been recovered naturally because the interaction energy $\rho_f \varphi$ was consistently included in the energy density functional. In the absence of free charges ($\rho_f = 0$), this reduces to the familiar source-free form $D_{i,i} = 0$, which is used in the subsequent plane-wave analysis.

\medskip
From \cref{eq:pre_traction,eq:pre_charge}, we obtain the natural boundary conditions on the surface $\partial V$:
\begin{align}
\sigma_{ij}^\alpha n_j &= t_i^\alpha, \label{eq:bc_traction} \\
D_i n_i &= \sigma_e. \label{eq:bc_charge}
\end{align}
The first condition states that the traction (stress vector) on the boundary equals the prescribed surface force $t_i^\alpha$. The second condition states that the normal component of the electric displacement on the boundary equals the prescribed surface charge density $\sigma_e$. These boundary conditions emerge naturally from the surface integrals in the localized energy balance, not as additional assumptions.

Together, \cref{eq:momentum,eq:gauss,eq:bc_traction,eq:bc_charge} form the complete set of field equations and boundary conditions for the linear multiphase porous piezoelectric continuum. They are derived entirely from the global energy conservation principle, without invoking Newton's second law or Hamilton's variational principle.

\subsubsection{Thermodynamic reciprocity of the electromechanical coupling coefficients: Maxwell relations}\label{sec:maxwell}

Once the potential energy density functional $\Psi(\varepsilon_{ij}^\alpha, E_i)$ has been specified, the constitutive relations of the multiphase piezoelectric medium are given by the definition of the thermodynamic conjugate forces. For the mechanical and electrical fields, the generalized forces are defined as
\begin{equation}
\sigma_{ij}^\alpha \equiv \frac{\partial \Psi}{\partial \varepsilon_{ij}^\alpha}, \qquad
D_i \equiv -\frac{\partial \Psi}{\partial E_i}.
\label{eq:conjugate_def}
\end{equation}
This definition is a standard result of continuum thermodynamics \cite{Malvern1969,Tiersten1969}.

It is worth emphasizing that the minus sign in the definition of the electric displacement $D_i$ is not an artificial convention; it originates from the Legendre transformation that changes the independent variable from the electric displacement $D_i$ to the electric field $E_i$.

\subsubsection{Thermodynamic differential of the electric enthalpy density}\label{sec:enthalpy_diff}

In the present theory, the material electric enthalpy density $\Psi$ is a function of two sets of state variables, the strain tensor $\varepsilon_{ij}^\alpha$ and the electric field $E_i$:
\begin{equation}
\Psi = \Psi(\varepsilon_{ij}^\alpha,\; E_i),
\label{eq:Psi_vars}
\end{equation}
whose explicit form is given in \cref{eq:psi_def}. Its conjugate variables are identified as
\begin{align}
\frac{\partial \Psi}{\partial \varepsilon_{ij}^\alpha} &= C_{ijkl}^{\alpha\beta} \varepsilon_{kl}^\beta - e_{kij}^\alpha E_k \equiv \sigma_{ij}^\alpha,
\label{eq:sigma_from_Psi} \\
\frac{\partial \Psi}{\partial E_i} &= - e_{ijk}^\alpha \varepsilon_{jk}^\alpha - \kappa_{ij} E_j \equiv -D_i,
\label{eq:D_from_Psi}
\end{align}
so that its total differential takes the standard two-term form
\begin{equation}
{\rm d}\Psi =
\sigma_{ij}^\alpha \,{\rm d}\varepsilon_{ij}^\alpha
- D_i \,{\rm d}E_i .
\label{eq:Psi_diff}
\end{equation}
This is the form commonly used in the variational derivation of piezoelectric field equations \cite{Tiersten1969, Maugin1988}.

Since $\Psi$ is a $C^2$ smooth function of the state variables, its second-order mixed partial derivatives are independent of the order of differentiation, and Schwarz's theorem \cite{Rudin1976} gives
\begin{equation}
\frac{\partial^2 \Psi}{\partial E_i \, \partial \varepsilon_{jk}^\alpha}
= \frac{\partial^2 \Psi}{\partial \varepsilon_{jk}^\alpha \, \partial E_i}.
\label{eq:schwarz}
\end{equation}

We now substitute the definitions of \cref{eq:conjugate_def} into the two sides of \cref{eq:schwarz}. For the left-hand side, differentiating first with respect to $\varepsilon$ and then with respect to $E$ gives
\begin{equation}
\frac{\partial}{\partial E_i} \left( \frac{\partial \Psi}{\partial \varepsilon_{jk}^\alpha} \right)
= \frac{\partial \sigma_{jk}^\alpha}{\partial E_i};
\label{eq:schwarz_lhs}
\end{equation}
for the right-hand side, differentiating first with respect to $E$ and then with respect to $\varepsilon$ gives
\begin{equation}
\frac{\partial}{\partial \varepsilon_{jk}^\alpha} \left( \frac{\partial \Psi}{\partial E_i} \right)
= \frac{\partial}{\partial \varepsilon_{jk}^\alpha} \left( -D_i \right)
= - \frac{\partial D_i}{\partial \varepsilon_{jk}^\alpha}.
\label{eq:schwarz_rhs}
\end{equation}

By Schwarz's theorem the left-hand side equals the right-hand side, and we immediately obtain the Maxwell reciprocity relation in multiphase piezoelectric media \cite{Munn1973}:
\begin{equation}
\left. \frac{\partial \sigma_{jk}^\alpha}{\partial E_i} \right|_{\varepsilon}
= - \left. \frac{\partial D_i}{\partial \varepsilon_{jk}^\alpha} \right|_E.
\label{eq:maxwell_general}
\end{equation}

Substituting the piezoelectric coefficients $e_{ijk}^\alpha$ already defined in \cref{eq:e_def} into the above relation, the explicit form of the Maxwell relation follows:
\begin{equation}
\left. \frac{\partial \sigma_{jk}^\alpha}{\partial E_i} \right|_{\varepsilon} = -e_{ijk}^\alpha, \qquad
\left. \frac{\partial D_i}{\partial \varepsilon_{jk}^\alpha} \right|_E = e_{ijk}^\alpha,
\label{eq:maxwell_explicit}
\end{equation}
or, equivalently, in the combined form
\begin{equation}
\left. \frac{\partial \sigma_{jk}^\alpha}{\partial E_i} \right|_{\varepsilon}
=- \left. \frac{\partial D_i}{\partial \varepsilon_{jk}^\alpha} \right|_E
= -e_{ijk}^\alpha.
\label{eq:maxwell_final}
\end{equation}

Physical meaning and self-consistency check: \cref{eq:maxwell_general,eq:maxwell_explicit,eq:maxwell_final} precisely guarantee the strict thermodynamic reciprocity between the ``direct piezoelectric effect'' (the response of the electric displacement to the mechanical strain) and the ``converse piezoelectric effect'' (the response of the stress to the electric field) \cite{Tiersten1969, Ciarletta1993}. Direct substitution of the constitutive equations derived below (see \cref{eq:constitutive_stress,eq:constitutive_D}) into \cref{eq:maxwell_final} shows that the two sides are identically equal. This demonstrates a Maxwell-type constitutive reciprocity that follows from the common potential and the equality of mixed partial derivatives. It should not be identified with the Onsager reciprocal relations, which concern linear flux--force relations in irreversible thermodynamics \cite{Onsager1931, Onsager1931b}.

\noindent\textbf{Remark:} The result of this subsection is general: the Maxwell reciprocity relation in \cref{eq:maxwell_final} applies not only to two-phase porous piezoelectric media but also to arbitrary multiphase ($M \ge 2$) situations. For a non-piezoelectric phase (such as a fluid phase), the corresponding piezoelectric coefficients satisfy $e_{ijk}^{\alpha} = 0$, and the above relation automatically reduces to the special case in which that phase does not participate in the electromechanical coupling.

\subsubsection{Multiphase constitutive equations}\label{sec:constitutive}

The potential energy density $\Psi(\varepsilon_{ij}^\alpha, E_i)$ derived from the energy conservation law, \cref{eq:psi_def}, is the starting point for constructing the constitutive relations of the piezoelectric medium. Differentiating $\Psi$ with respect to the state variables, namely the strain $\varepsilon_{ij}^\alpha$ and the electric field $E_i$, directly yields the constitutive equations for the stress and the electric displacement:
\begin{equation}
\sigma_{ij}^\alpha = \frac{\partial \Psi}{\partial \varepsilon_{ij}^\alpha} = C_{ijkl}^{\alpha\beta} \varepsilon_{kl}^\beta - e_{kij}^\alpha E_k,
\label{eq:constitutive_stress}
\end{equation}
\begin{equation}
D_i = -\frac{\partial \Psi}{\partial E_i} = e_{ijk}^\alpha \varepsilon_{jk}^\alpha + \kappa_{ij} E_j.
\label{eq:constitutive_D}
\end{equation}
This derivation involves only partial differentiation and introduces no additional physical assumption. The forms of \cref{eq:constitutive_stress,eq:constitutive_D} are consistent with the standard constitutive equations of Tiersten's linear piezoelectric theory \cite{Tiersten1969, Maugin1988}, which shows that the constitutive relations obtained from energy conservation are equivalent to those of the traditional thermodynamic method and confirms the consistency of the present approach.

In particular, \cref{eq:constitutive_stress,eq:constitutive_D} both originate from the same potential function $\Psi$. According to Schwarz's theorem (the symmetry of mixed partial derivatives, \cref{eq:schwarz}), if $\Psi$ is sufficiently smooth with respect to $\varepsilon_{ij}^\alpha$ and $E_i$, the second-order mixed partial derivatives are independent of the order of differentiation:
\begin{equation}
\frac{\partial^2 \Psi}{\partial E_i \partial \varepsilon_{jk}^\alpha}
= \frac{\partial^2 \Psi}{\partial \varepsilon_{jk}^\alpha \partial E_i}.
\label{eq:schwarz_repeat}
\end{equation}
This is precisely the Maxwell reciprocity relation of \cref{eq:maxwell_final}; the corresponding symmetries of the elastic and dielectric constants, $C_{ijkl}^{\alpha\beta}=C_{klij}^{\beta\alpha}$ and $\kappa_{ij}=\kappa_{ji}$, follow in the same way from Schwarz's theorem applied to derivatives with respect to like variables. The symmetry
\begin{equation}
e_{ijk}^\alpha = e_{ikj}^\alpha,
\label{eq:e_sym}
\end{equation}
has a different origin: it follows from the symmetry of the strain tensor. Therefore, this constitutive model is not obtained by artificially patching together empirical formulas; it is generated naturally from the exact differential structure of a single energy function, which guarantees the thermodynamic self-consistency of the theory.

The consistency and thermodynamic robustness of the governing equations and constitutive relations derived from the energy-conservation framework are independently confirmed by re-deriving the same set of equations from Hamilton's variational principle; the detailed derivation is given in \cref{sec:validation_HP}.

\section{Linear plane-wave propagation}\label{sec:planewave}

\subsection{Plane-wave assumption and generalized Christoffel equation}\label{sec:christoffel}

Neglecting body forces ($f_i^\alpha = 0$) and free charges ($\rho_f = 0$), consider a harmonic plane wave propagating along the unit vector $\bm{n}$ ($n_j n_j = 1$), with angular frequency $\omega$, wavenumber $k$, and phase velocity $v = \omega/k$. The displacements of each phase and the electric potential are taken as
\begin{equation}
u_i^\alpha = U_i^\alpha \exp[{\rm i} k (n_j x_j - v t)], \qquad \varphi = \Phi \exp[{\rm i} k (n_j x_j - v t)].
\label{eq:plane_wave}
\end{equation}
Substituting the plane-wave solutions \cref{eq:plane_wave}, together with the geometric relations, into the momentum equation \cref{eq:momentum} and the constitutive equation \cref{eq:constitutive_stress}, and cancelling the common exponential factor and $k^2$, we obtain
\begin{equation}
\left( C_{ijkl}^{\alpha\beta} n_j n_k - \rho_{il}^{\alpha\beta} v^2 \right) U_l^\beta + e_{kij}^\alpha n_j n_k \Phi = 0.
\label{eq:christoffel_mech}
\end{equation}
Similarly, substitution into Gauss's law \cref{eq:gauss} and the constitutive equation \cref{eq:constitutive_D} yields
\begin{equation}
e_{ijk}^\alpha n_i n_j U_k^\alpha - \kappa_{ij} n_i n_j \Phi = 0.
\label{eq:christoffel_elec}
\end{equation}
We define the \textbf{multiphase Christoffel tensor} $\Gamma_{il}^{\alpha\beta} = C_{ijkl}^{\alpha\beta} n_j n_k$, the \textbf{piezoelectric coupling vector} $\gamma_i^\alpha = e_{kij}^\alpha n_j n_k$, and the \textbf{effective permittivity} $\epsilon = \kappa_{ij} n_i n_j$.

\subsection{Matrix eigenvalue problem}\label{sec:eigen}

Combining the above equations and assembling the state vector $\bm{X} = [U_1^1, U_2^1, U_3^1, \dots, U_1^M, U_2^M, U_3^M, \Phi]^T$ of dimension $(3M+1) \times 1$, the system becomes the generalized eigenvalue problem
\begin{equation}
\left( \bm{K} - v^2 \bm{M} \right) \bm{X} = \bm{0},
\label{eq:eigenvalue}
\end{equation}
where the generalized stiffness matrix $\bm{K}$ and the generalized mass matrix $\bm{M}$ have the block structures
\begin{equation}
\bm{K} = \begin{bmatrix}
\bm{\Gamma}^{11} & \cdots & \bm{\Gamma}^{1M} & \bm{\gamma}^1 \\
\vdots & \ddots & \vdots & \vdots \\
\bm{\Gamma}^{M1} & \cdots & \bm{\Gamma}^{MM} & \bm{\gamma}^M \\
(\bm{\gamma}^1)^T & \cdots & (\bm{\gamma}^M)^T & -\epsilon
\end{bmatrix}, \quad
\bm{M} = \begin{bmatrix}
\bm{\rho}^{11} & \cdots & \bm{\rho}^{1M} & \bm{0} \\
\vdots & \ddots & \vdots & \vdots \\
\bm{\rho}^{M1} & \cdots & \bm{\rho}^{MM} & \bm{0} \\
\bm{0}^T & \cdots & \bm{0}^T & 0
\end{bmatrix}.
\label{eq:matrices}
\end{equation}

\subsection{Properties of the generalized eigenvalue system}\label{sec:eigen_props}

\begin{enumerate}[leftmargin=*]
  \item \textbf{Number of wave modes}: the matrix pencil $(\bm{K},\bm{M})$ has dimension $3M+1$, but the last row and column of $\bm{M}$ vanish because the electric potential carries no inertia. The potential therefore acts as a quasi-static constraint rather than as an independent propagating mode: eliminating $\Phi$ through the last row of \cref{eq:eigenvalue} yields the condensed Christoffel tensor $\bm{\Gamma}_{\rm eff} = \bm{\Gamma} + \bm{\gamma}\bm{\gamma}^{T}/\epsilon$, in which the piezoelectric term stiffens the mechanical system. The characteristic equation $\det(\bm{K} - v^2 \bm{M}) = 0$ is consequently of degree $3M$ in $v^2$, and the number of finite propagating modes is determined by the rank of the mechanical subsystem; an inviscid fluid phase, which possesses no shear stiffness, contributes zero roots.
  \item \textbf{Thermodynamic self-consistency}: the symmetry of the Hessian matrix of the potential energy $\Psi$ implies that the matrix $\bm{K}$ is strictly symmetric, which proves that the multiphase coupled constitutive relations satisfy the Maxwell reciprocity relations.
  \item \textbf{Interphase coupling mechanism}: the off-diagonal blocks $\bm{\rho}^{\alpha\beta}$ of the mass matrix $\bm{M}$ naturally describe the interphase inertial coupling effects.
\end{enumerate}

\subsection{Reduction to a two-phase porous piezoelectric medium}\label{sec:twophase}

As a first test of the self-consistency of the theory, we now reduce the above multiphase framework to an important limiting case: the two-phase porous piezoelectric medium. Take the number of phases $M=2$, and denote the solid skeleton phase by $\alpha=s$ and the pore fluid phase by $\alpha=f$. Since piezoelectricity originates only from the noncentrosymmetric crystal structure of the solid skeleton, while the fluid phase (liquid or gas) possesses no piezoelectricity, the piezoelectric stress tensors satisfy
\begin{equation}
e_{ijk}^{f}=0,\qquad e_{ijk}^{s}\neq 0 .
\label{eq:e_twophase}
\end{equation}

In this case, the potential energy density functional \cref{eq:psi_def} reduces to
\begin{equation}
\begin{aligned}
\Psi =\,& \frac12 C_{ijkl}^{ss}\varepsilon_{ij}^{s}\varepsilon_{kl}^{s}
+C_{ijkl}^{sf}\varepsilon_{ij}^{s}\varepsilon_{kl}^{f}
+\frac12 C_{ijkl}^{ff}\varepsilon_{ij}^{f}\varepsilon_{kl}^{f} \\
&- e_{ijk}^{s}E_i\varepsilon_{jk}^{s}
-\frac12\kappa_{ij}E_iE_j ,
\end{aligned}
\label{eq:psi_twophase}
\end{equation}
where $C_{ijkl}^{ss}$ and $C_{ijkl}^{ff}$ are the self-coupling elastic tensors of the solid skeleton and of the fluid, respectively, and $C_{ijkl}^{sf}$ describes the solid--fluid interphase elastic coupling; physically it reflects the contribution of the pore pressure to the deformation of the skeleton, and corresponds to the coupling modulus of Biot's theory \cite{Biot1956}. The transposed cross block follows from the symmetry of the Hessian, $C_{ijkl}^{fs}=C_{klij}^{sf}$.

The corresponding constitutive relations follow directly from \cref{eq:constitutive_stress,eq:constitutive_D}:
\begin{align}
\sigma_{ij}^{s} &= C_{ijkl}^{ss}\varepsilon_{kl}^{s}+C_{ijkl}^{sf}\varepsilon_{kl}^{f}-e_{kij}^{s}E_k,\label{eq:sigma_s}\\
\sigma_{ij}^{f} &= C_{ijkl}^{fs}\varepsilon_{kl}^{s}+C_{ijkl}^{ff}\varepsilon_{kl}^{f},\label{eq:sigma_f}\\
D_i &= e_{ijk}^{s}\varepsilon_{jk}^{s}+\kappa_{ij}E_j .\label{eq:D_twophase}
\end{align}

The momentum equation \cref{eq:momentum} gives the dynamic equations of the solid and fluid phases, respectively:
\begin{align}
\rho_{ij}^{ss}\ddot u_j^{s}+\rho_{ij}^{sf}\ddot u_j^{f}
&=\sigma_{ij,j}^{s}+f_i^{s},\label{eq:mom_s}\\
\rho_{ij}^{fs}\ddot u_j^{s}+\rho_{ij}^{ff}\ddot u_j^{f}
&=\sigma_{ij,j}^{f}+f_i^{f},\label{eq:mom_f}
\end{align}
where the off-diagonal density tensor $\rho_{ij}^{sf}$ naturally characterizes the solid--fluid interphase inertial coupling effect, without introducing any phenomenological ``virtual mass'' parameter. This feature is physically consistent with the treatment of interphase inertia through kinetic-energy coupling coefficients in Biot's theory; here, however, it is derived directly from the second-order expansion coefficients of the energy density functional, and thus has a first-principles basis \cite{Biot1956,Coussy2004}.

For the linear plane-wave propagation problem, the generalized Christoffel equation \cref{eq:eigenvalue} reduces to a $7\times7$ ($3\times 2 + 1$) eigenvalue problem, whose generalized stiffness matrix $\bm{K}$ and mass matrix $\bm{M}$ have the block structures
\begin{equation}
\bm{K}=
\begin{bmatrix}
\bm{\Gamma}^{ss} & \bm{\Gamma}^{sf} & \bm{\gamma}^{s}\\
\bm{\Gamma}^{fs} & \bm{\Gamma}^{ff} & \bm{0}\\
(\bm{\gamma}^{s})^{T} & \bm{0}^{T} & -\epsilon
\end{bmatrix},\qquad
\bm{M}=
\begin{bmatrix}
\bm{\rho}^{ss} & \bm{\rho}^{sf} & \bm{0}\\
\bm{\rho}^{fs} & \bm{\rho}^{ff} & \bm{0}\\
\bm{0}^{T} & \bm{0}^{T} & 0
\end{bmatrix},
\label{eq:KM_twophase}
\end{equation}
where $\bm{\gamma}_i^{s}=e_{kij}^{s} n_j n_k$ depends only on the piezoelectric tensor of the solid skeleton. The piezoelectric coupling is embodied entirely in the cross term between the solid-phase strain and the electric field; the fluid phase, having $e_{ijk}^{f}=0$, does not participate directly in the electromechanical coupling, and is affected by the piezoelectric effect only indirectly through the solid--fluid mechanical coupling.

The above solid- and fluid-phase constitutive relations (the two-phase reduced forms of \cref{eq:constitutive_stress,eq:constitutive_D}) and the coupled momentum equations are structurally consistent with the classical theories of poromechanics and of piezoelectric continua; the reduced form is compatible with the thermodynamic formulations developed by Ciarletta and co-workers \cite{Ciarletta1993,Ciarletta1996} and with Biot--Tiersten-type porous piezoelectric models. Specifically, when the piezoelectric effect is neglected ($e_{ijk}^{s}=0$), the above system of equations reduces exactly to Biot's wave equations for saturated porous media \cite{Biot1956}; when, instead, the fluid phase is neglected ($M=1$) while the solid-phase piezoelectric coupling is retained, the system reduces to Tiersten's single-phase piezoelectric theory \cite{Tiersten1969}. More importantly, in the present framework all coupling coefficients originate from the same potential function $\Psi$, and Schwarz's theorem automatically guarantees that the Maxwell reciprocity relation holds, namely
\begin{equation}
\left.\frac{\partial \sigma_{jk}^{s}}{\partial E_i}\right|_{\varepsilon}
=
-\left.\frac{\partial D_i}{\partial \varepsilon_{jk}^{s}}\right|_E
=
-e_{ijk}^{s},
\label{eq:maxwell_twophase}
\end{equation}
which establishes the corresponding Maxwell reciprocity relation within the present constitutive formulation.

In summary, the multiphase energy-conservation framework proposed in this paper is a natural unified generalization of Biot's porous-media theory and of Tiersten's single-phase piezoelectric theory, and is compatible with the thermodynamic formulations of porous piezoelectric materials developed by Ciarletta and co-workers. In the corresponding limits the governing equations of Biot's and Tiersten's theories are recovered term by term; a numerical illustration of the corresponding modal structures is given in \cref{sec:numerical}.

To display intuitively the parameter mapping between the present theory and the classical theories, \cref{tab:mapping} summarizes the correspondence between the key material parameters of the two-phase porous piezoelectric model of this paper and the corresponding physical quantities of Biot's porous-media theory and Tiersten's single-phase piezoelectric theory.

\begin{table*}[t]
\footnotesize
\begin{threeparttable}
\caption{Correspondence between the parameters of the present theory and those of the Biot and Tiersten theories}
\label{tab:mapping}
\doublerulesep 0.1pt \tabcolsep 4pt
\begin{tabular*}{\textwidth}{@{\extracolsep{\fill}}>{\raggedright\arraybackslash}p{0.40\textwidth}>{\raggedright\arraybackslash}p{0.28\textwidth}>{\raggedright\arraybackslash}p{0.20\textwidth}}
\toprule
Present parameter (two-phase porous piezoelectric) & Biot theory counterpart & Tiersten theory counterpart \\\hline
$\rho_{ij}^{ss}$ (solid self-coupling density) & $\rho_{11}\delta_{ij}$ (generalized mass coefficient) & $\rho\delta_{ij}$ (mass density) \\
$\rho_{ij}^{ff}$ (fluid self-coupling density) & $\rho_{22}\delta_{ij}$ (generalized mass coefficient) & --- \\
$\rho_{ij}^{sf}=\rho_{ji}^{fs}$ (solid--fluid inertial coupling density) & $\rho_{12}\delta_{ij}$ (inertial coupling coefficient) & --- \\
$C_{ijkl}^{ss}$ (solid self-coupling stiffness) & $P_{ijkl}$ (solid-block stiffness) & $c_{ijkl}$ (elastic stiffness) \\
$C_{ijkl}^{ff}$ (fluid self-coupling stiffness) & $R\, \delta_{ij}\delta_{kl}$ (fluid-block modulus) & --- \\
$C_{ijkl}^{sf}$ (solid--fluid elastic coupling) & $Q\, \delta_{ij}\delta_{kl}$ (solid--fluid coupling modulus) & --- \\
$e_{ijk}^{s}$ (solid-phase piezoelectric stress tensor) & --- & $e_{ijk}$ (piezoelectric stress constants) \\
$\kappa_{ij}$ (dielectric tensor) & --- & $\kappa_{ij}$ (dielectric constant tensor) \\
\bottomrule
\end{tabular*}
\end{threeparttable}
\end{table*}

\Cref{tab:mapping} is written in Biot's two-displacement $(u,U)$ form, in which the phase stresses are related to the phase strains through the stiffness blocks $P$, $Q$, and $R$, and the kinetic energy involves the generalized mass coefficients $\rho_{11}$, $\rho_{12}$, and $\rho_{22}$ \cite{Biot1956}. In this form the solid block $C_{ijkl}^{ss}$ is the stiffness of the fluid-loaded solid phase and should not be confused with the drained (dry-frame) stiffness; likewise, $\rho_{11}$ and $\rho_{22}$ are not the intrinsic solid and fluid densities, but combine the phase densities with the added (virtual) mass associated with the relative solid--fluid motion.

It should be noted that the correspondence in \cref{tab:mapping} holds in the following limiting senses.

\begin{enumerate}[leftmargin=*]
\item \textbf{Reduction to Biot's theory}: when the piezoelectric effect vanishes ($e_{ijk}^{s}=0$), the solid--fluid two-phase model of this paper reduces exactly to Biot's theory of saturated porous media \cite{Biot1956,Biot1956b}. In this limit, the inertial coupling density tensor $\rho_{ij}^{sf}$ naturally replaces the ``virtual mass'' parameter $\rho_a$ (or $\rho_{12}$) that has to be introduced separately in Biot's theory, and possesses a well-defined first-principles definition without any phenomenological correction. The elastic coupling tensor $C_{ijkl}^{sf}$ corresponds to the coupling modulus $Q$ of Biot's $(u,U)$ formulation, which describes the coupling between the pore pressure and the volumetric strain of the skeleton.

\item \textbf{Reduction to Tiersten's theory}: when the fluid phase is neglected ($M=1$, $C_{ijkl}^{sf}=0$, $\rho_{ij}^{sf}=0$), the present model reduces to the single-phase piezoelectric continuum theory. In this limit, the solid self-coupling elastic tensor $C_{ijkl}^{ss}$ reduces to the elastic stiffness tensor $c_{ijkl}$ of Tiersten's theory, and the piezoelectric stress tensor $e_{ijk}^{s}$ and the dielectric tensor $\kappa_{ij}$ correspond to the piezoelectric and dielectric constants of Tiersten's theory, respectively \cite{Tiersten1969}.

\item \textbf{Relation to earlier thermodynamic theories of porous piezoelectric materials}: when the piezoelectric effect is included and the solid--fluid two-phase coupling is retained, the constitutive relations of this paper are compatible with the thermodynamic restrictions on the constitutive equations of porous piezoelectric materials established by Ciarletta and co-workers within a linear thermodynamic framework \cite{Ciarletta1993,Ciarletta1996}. Compared with these classical works, the advantage of the present theory is that all parameters are derived from a unified energy density functional, so that the Maxwell reciprocity relation ($\partial\sigma_{jk}^{s}/\partial E_i = -\,\partial D_i/\partial\varepsilon_{jk}^{s}$) is satisfied automatically, without imposing additional constitutive restrictions.
\end{enumerate}

The parameter mapping relations above clearly show that the energy-conservation framework for multiphase porous piezoelectric media proposed in this paper incorporates Biot's porous-media theory and Tiersten's piezoelectric theory into a single self-consistent first-principles system, in which each of these classical theories is recovered naturally as a special case, and that it is compatible with the thermodynamic formulations of porous piezoelectric materials developed by Ciarletta and co-workers.

\subsection{Numerical illustration of representative limiting cases}\label{sec:numerical}

To display the modal structures contained in the unified equations, we compute the phase velocities of three representative limiting models: (A) the Biot two-phase poroelastic limit, obtained from the two-phase equations by setting $e_{ijk}^{s}=0$; (B) the Tiersten-type single-phase piezoelectric limit, assembled independently with the intrinsic constants of dense PZT-2; and (C) the full two-phase porous piezoelectric model. The materials are taken from the same PZT-2 family: the solid skeleton is PZT-2, a transversely isotropic piezoelectric ceramic poled along the $z$ axis, and in the two-phase models the pore space is filled with water. Models A and C share the same two-phase parameter set and differ only in the piezoelectric coupling, so their comparison isolates the electromechanical effect. Model B, in contrast, is a dense single-phase reference; its velocities differ from those of models A and C through the porosity, the effective stiffness, and the density as well, and the differences should not be attributed to piezoelectricity alone. All three models are dissipationless, consistent with the time-reversal symmetry assumed in \cref{sec:axioms}. The electric potential is eliminated through the quasi-static Gauss constraint, \cref{eq:christoffel_elec}, so that the velocities follow from a real symmetric generalized eigenvalue problem. We consider plane waves in the $x$--$z$ plane, with the propagation angle $\theta$ measured from the poling axis $z$.

The material constants are listed in \cref{tab:params} and follow the notation of the general theory, with the piezoelectric constants written in Voigt notation. For the two-phase medium, $C_{11}$, $C_{13}$, $C_{33}$, and $C_{44}$ are the Voigt components of the solid-block tensor $C_{ijkl}^{ss}$ in the phase-stress form of \cref{tab:mapping}; $m_{11}$ and $m_{33}$ are the two independent normal components of the coupling block $C_{ijkl}^{sf}$; and $R$ is the fluid modulus, $C_{ijkl}^{ff} = R\,\delta_{ij}\delta_{kl}$. The density tensors are taken to be isotropic, $\rho_{ij}^{\alpha\beta} = \rho^{\alpha\beta}\delta_{ij}$, that is, the inertial anisotropy of the pore space is neglected. Models A and C share all two-phase parameters, and model A sets $e_{15}=e_{31}=e_{33}=0$. For model B, the same symbols denote the intrinsic constants $c_{ijkl}$, $e_{ijk}$, and $\kappa_{ij}$ of dense PZT-2.

\begin{table*}[t]
\footnotesize
\begin{threeparttable}
\caption{Material constants used in the numerical illustration (models A and C: water-saturated porous PZT-2; model B: dense PZT-2)}
\label{tab:params}
\doublerulesep 0.1pt \tabcolsep 3pt
\begin{tabular*}{\textwidth}{@{\extracolsep{\fill}}lcc}
\toprule
Parameter group & Two-phase porous medium (A, C) & Dense PZT-2 (B) \\\hline
$C_{11},\,C_{13},\,C_{33},\,C_{44}$ (GPa) & 83.83,\ 32.26,\ 71.39,\ 15.82 & 148.0,\ 74.2,\ 131.0,\ 25.3 \\
$m_{11},\,m_{33},\,R$ (GPa) & 0.638,\ 0.661,\ 0.426 & --- \\
$\rho^{ss},\,\rho^{sf},\,\rho^{ff}$ (kg m$^{-3}$) & 6180,\ $-100$,\ 300 & $\rho=7600$ \\
$e_{15},\,e_{31},\,e_{33}$ (C m$^{-2}$) & 5.525,\ $-0.583$,\ 6.402 & 9.8,\ $-1.9$,\ 9.0 \\
$\kappa_{11},\,\kappa_{33}$ ($10^{-9}$ F m$^{-1}$) & 3.579,\ 1.722 & 4.463,\ 2.302 \\
\bottomrule
\end{tabular*}
\end{threeparttable}
\end{table*}

\begin{figure*}[t]
\centering
\includegraphics[width=0.98\textwidth]{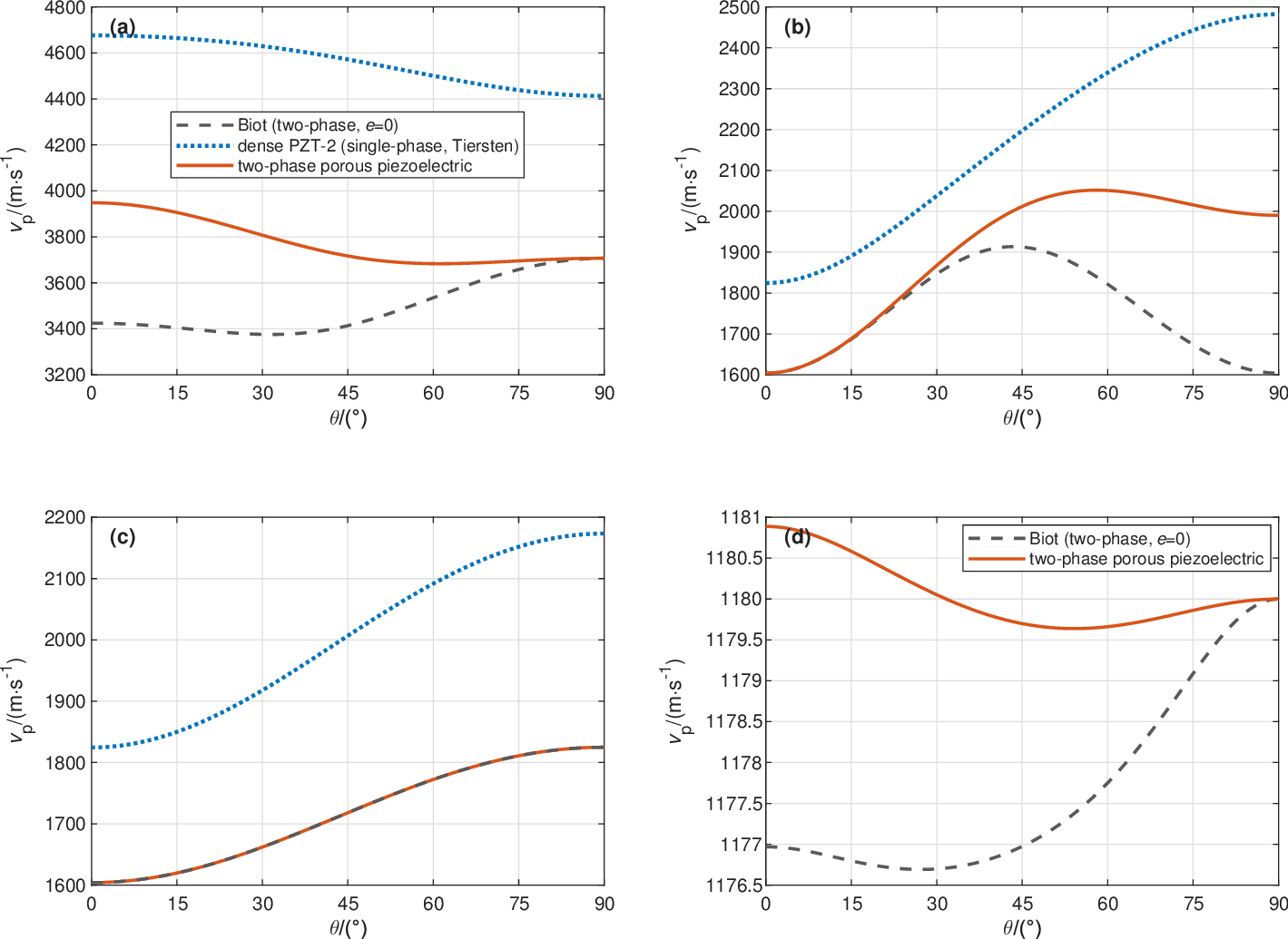}
\caption{Phase velocities of the three limiting models versus the propagation angle $\theta$ measured from the poling axis: (a) fast compressional modes, (b) quasi-shear (qSV) modes, (c) SH modes, and (d) the slow compressional mode qP2 on an expanded vertical scale. Dashed curves, Biot two-phase model (A); dotted curves, dense single-phase PZT-2 (B); solid curves, two-phase porous piezoelectric model (C).}
\label{fig:velocity}
\end{figure*}

\Cref{fig:velocity} shows the phase velocities of the three models as functions of $\theta$. Four features of the modal structures deserve attention. First, the single-phase piezoelectric limit supports a quasi-compressional, a quasi-shear, and an SH branch but no counterpart of the Biot slow wave, because it contains no independent pore-fluid degree of freedom. Second, the Biot and porous piezoelectric curves coincide in symmetry-selected directions: the fast compressional branches coincide at $\theta=90^\circ$, and the quasi-shear branches coincide at $\theta=0^\circ$. In these directions the relevant projection of the piezoelectric coupling vector $\bm{\gamma}^{s}$ vanishes, so the electromechanical stiffening disappears; the influence of the coupling on the wave velocities is therefore strongly direction dependent. Third, for the SH branch, whose displacement is polarized normal to the $x$--$z$ propagation plane, the piezoelectric coupling projection vanishes identically; the Biot and porous piezoelectric SH curves therefore coincide over the entire angular range, while the pore fluid affects this branch only through the inertial coupling. Fourth, the slow compressional branch qP2, drawn on an expanded vertical scale in panel (d), is only weakly shifted by the piezoelectric coupling, which reaches this fluid-borne mode indirectly through the solid--fluid coupling; its existence is controlled primarily by the pore-fluid degree of freedom.

\section{Conclusions}\label{sec:conclusions}

In this paper, a linear theory of elastic-wave dynamics in multiphase porous piezoelectric media has been constructed on the basis of global energy conservation. The main conclusions are as follows.
\begin{enumerate}[leftmargin=*]
  \item \textbf{Expansion of the energy function}: starting from the general formula of the multivariate Taylor expansion, the energy density of the multiphase system was expanded to second order about the natural equilibrium state. The natural-equilibrium condition eliminates the first-order terms, and time-reversal symmetry shows that the cross terms containing an odd number of velocities must vanish.
  \item \textbf{Symmetry and thermodynamic constraints of the electromechanical coupling}: the mathematical basis for the existence of the electromechanical coupling term was analysed. Time-reversal symmetry permits its existence (a product of even functions); spatial inversion symmetry determines whether it is nonzero (the material must be noncentrosymmetric); and Schwarz's theorem guarantees the Maxwell reciprocity between the direct and converse piezoelectric effects, ensuring the self-consistency of the theory.
  \item \textbf{Emergence of the dynamic equations and wave propagation characteristics}: by means of the localization of the energy conservation integral and the arbitrariness of the test functions, the multiphase momentum equations, Gauss's law, and the constitutive equations emerge naturally as mathematical corollaries. The generalized Christoffel eigenvalue equation derived for the linear plane-wave problem clearly reveals the mechanical wave branches of multiphase media, with the electric potential entering as a quasi-static constraint that stiffens the mechanical system rather than as an additional propagating mode.
  \item \textbf{Numerical illustration of the limiting cases}: phase-velocity calculations for three representative limiting models illustrate the recovery of the Biot and Tiersten modal structures within the unified equations, and show that the influence of the electromechanical coupling on the wave velocities is strongly direction dependent in the two-phase porous piezoelectric case.
\end{enumerate}

The present theoretical framework is mathematically rigorous and physically transparent. It provides a solid first-principles foundation for the acoustic characterization of multiphase porous piezoelectric media and for the design of piezoelectric metamaterials.

\section*{CRediT authorship contribution statement}

\textbf{Xiuming Wang:} Conceptualization, Methodology, Writing -- original
draft, Supervision.
\textbf{Yinqiu Zhou:} Validation, Supervision, Writing -- review and editing.
\textbf{Zhixiang Sun:} Software, Formal analysis, Validation, Visualization,
Writing -- review and editing.
\textbf{Lin Liu:} Validation, Writing -- review and editing.

\section*{Funding}

This work was supported by the National Natural Science Foundation of China
[grant numbers 12274432, 52227901, 42127807].


\begin{thebibliography}{99}
\expandafter\ifx\csname url\endcsname\relax
  \def\url#1{\texttt{#1}}\fi
\expandafter\ifx\csname urlprefix\endcsname\relax\def\urlprefix{URL }\fi
\expandafter\ifx\csname href\endcsname\relax
  \def\href#1#2{#2} \def\path#1{#1}\fi

\bibitem{Rohan2018}
E.~Rohan, V.~Lukeš, Homogenization of the fluid-saturated piezoelectric porous
  media, Int. J. Solids Struct. 147 (2018) 110--125.
\newblock \href {https://doi.org/10.1016/j.ijsolstr.2018.05.017}
  {\path{doi:10.1016/j.ijsolstr.2018.05.017}}.

\bibitem{Tariq2025}
M.~H. Tariq, Y.-T. Zhou, Advanced 3{D} solutions for coupled
  mechanical--electrical interactions in porous-piezoelectric materials
  unraveling solid cone dynamics, Z. Angew. Math. Phys. 76 (2025) 144.
\newblock \href {https://doi.org/10.1007/s00033-025-02521-x}
  {\path{doi:10.1007/s00033-025-02521-x}}.

\bibitem{Kudimova2017}
A.~Kudimova, I.~Mikhayluts, D.~Nadolin, A.~Nasedkin, A.~Nasedkina,
  P.~Oganesyan, A.~Soloviev, Computer design of porous and ceramic
  piezocomposites in the finite element package {ACELAN}, Procedia Struct.
  Integrity 6 (2017) 301--308.
\newblock \href {https://doi.org/10.1016/j.prostr.2017.11.046}
  {\path{doi:10.1016/j.prostr.2017.11.046}}.

\bibitem{Arai1991}
T.~Arai, K.~Ayusawa, H.~Sato, T.~Miyata, K.~Kawamura, K.~Kobayashi, Properties
  of hydrophone with porous piezoelectric ceramics, Jpn. J. Appl. Phys. 30~(9S)
  (1991) 2253.
\newblock \href {https://doi.org/10.1143/jjap.30.2253}
  {\path{doi:10.1143/jjap.30.2253}}.

\bibitem{Kar2007}
R.~Kar-Gupta, T.~A. Venkatesh, Electromechanical response of porous
  piezoelectric materials: {E}ffects of porosity distribution, Appl. Phys.
  Lett. 91~(6) (2007) 062904.
\newblock \href {https://doi.org/10.1063/1.2766960}
  {\path{doi:10.1063/1.2766960}}.

\bibitem{Bowen2022}
Z.~Rymansaib, P.~Kurt, Y.~Zhang, J.~I. Roscow, C.~R. Bowen, A.~J. Hunter,
  Ultrasonic transducers made from freeze-cast porous piezoceramics, IEEE
  Trans. Ultrason. Ferroelectr. Freq. Control 69~(3) (2022) 1100--1111.
\newblock \href {https://doi.org/10.1109/TUFFC.2022.3144521}
  {\path{doi:10.1109/TUFFC.2022.3144521}}.

\bibitem{Tezcan2025}
A.~Tezcan, Processing of smart porous electro-ceramic transducers ({ProSPECT}),
  Dissertation for the {D}octoral {D}egree, University of Bath, Bath (2025).

\bibitem{Biot1956}
M.~A. Biot, Theory of propagation of elastic waves in a fluid-filled porous
  solid. {I}. {L}ow-frequency range, J. Acoust. Soc. Am. 28~(2) (1956)
  168--178.
\newblock \href {https://doi.org/10.1121/1.1908239}
  {\path{doi:10.1121/1.1908239}}.

\bibitem{Coussy2004}
O.~Coussy, Poromechanics, John Wiley \& Sons, Chichester, UK, 2004.

\bibitem{Santos1990}
J.~E. Santos, J.~Douglas, J.~Corbero, O.~M. Lovera, A model for wave
  propagation in a porous medium saturated by a two-phase fluid, J. Acoust.
  Soc. Am. 87~(4) (1990) 1439--1448.
\newblock \href {https://doi.org/10.1121/1.399440}
  {\path{doi:10.1121/1.399440}}.

\bibitem{Leclaire1994}
P.~Leclaire, F.~Cohen-T{\'e}noudji, J.~Aguirre-Puente, Extension of biot's
  theory of wave propagation to frozen porous media, J. Acoust. Soc. Am. 96~(6)
  (1994) 3753--3768.
\newblock \href {https://doi.org/10.1121/1.411336}
  {\path{doi:10.1121/1.411336}}.

\bibitem{Leclaire1995}
P.~Leclaire, F.~Cohen-Ténoudji, J.~Aguirre-Puente, Observation of two
  longitudinal and two transverse waves in a frozen porous medium, J. Acoust.
  Soc. Am. 97~(4) (1995) 2052--2055.
\newblock \href {https://doi.org/10.1121/1.411997}
  {\path{doi:10.1121/1.411997}}.

\bibitem{Tiersten1969}
H.~F. Tiersten, Linear Piezoelectric Plate Vibrations, Plenum Press, New York,
  1969.

\bibitem{Truesdell1960}
C.~Truesdell, R.~A. Toupin, The classical field theories, in: Principles of
  Classical Mechanics and Field Theory, Springer, Berlin, 1960, pp. 226--858.

\bibitem{Gurtin1972}
M.~E. Gurtin, The linear theory of elasticity, in: Mechanics of Solids {II},
  Springer, Berlin, 1972, pp. 1--295.

\bibitem{Biot1956b}
M.~A. Biot, Theory of propagation of elastic waves in a fluid-filled porous
  solid. {II}. {H}igher-frequency range, J. Acoust. Soc. Am. 28~(2) (1956)
  179--191.
\newblock \href {https://doi.org/10.1121/1.1908241}
  {\path{doi:10.1121/1.1908241}}.

\bibitem{Bedford1984}
A.~Bedford, R.~D. Costley, M.~Stern, On the drag and virtual mass coefficients
  in {Biot}'s equations, J. Acoust. Soc. Am. 76~(6) (1984) 1804--1809.
\newblock \href {https://doi.org/10.1121/1.391577}
  {\path{doi:10.1121/1.391577}}.

\bibitem{Capriz1987}
G.~Capriz, P.~Giovine, On effects of virtual inertia during diffusion of a
  dispersed medium in a suspension, Arch. Ration. Mech. Anal. 98~(2) (1987)
  115--122.

\bibitem{Yavari1988}
B.~Yavari, A.~Bedford, Computation of the {B}iot drag and virtual mass
  coefficients, Int. J. Multiphase Flow 14~(1) (1988) 1--12.
\newblock \href {https://doi.org/10.1016/0301-9322(88)90030-4}
  {\path{doi:10.1016/0301-9322(88)90030-4}}.

\bibitem{Cassel2013}
K.~W. Cassel, Variational Methods with Applications in Science and Engineering,
  Cambridge University Press, Cambridge, UK, 2013.

\bibitem{Zhou2022a}
Y.~Zhou, X.~Wang, A methodology for formulating dynamical equations in
  analytical mechanics based on the principle of energy conservation, J. Phys.
  Commun. 6~(3) (2022) 035006.
\newblock \href {https://doi.org/10.1088/2399-6528/ac57f8}
  {\path{doi:10.1088/2399-6528/ac57f8}}.

\bibitem{Wang2023}
X.-M. Wang, Y.-Q. Zhou, Research on elastodynamic theory based on the framework
  of energy conservation, Acta Phys. Sin. 72~(7) (2023) 074501.
\newblock \href {https://doi.org/10.7498/aps.72.20212272}
  {\path{doi:10.7498/aps.72.20212272}}.

\bibitem{Zhou2022b}
Y.~{Zhou}, X.~{Zhang}, L.~{Liu}, T.~{Liu}, X.~{Wang}, Formulations of the
  elastodynamic equations in anisotropic and multiphasic porous media from the
  principle of energy conservation, Prog. Theor. Exp. Phys. 2022~(12) (2022)
  123A01.
\newblock \href {https://doi.org/10.1093/ptep/ptac149}
  {\path{doi:10.1093/ptep/ptac149}}.

\bibitem{Ciarletta1993}
M.~Ciarletta, A.~Scalia, Thermodynamic theory for porous piezoelectric
  materials, Meccanica 28~(4) (1993) 303--308.
\newblock \href {https://doi.org/10.1007/BF00987166}
  {\path{doi:10.1007/BF00987166}}.

\bibitem{Ciarletta1996}
M.~Ciarletta, E.~Scarpetta, Some results on thermoelasticity for porous
  piezoelectric materials, Mech. Res. Commun. 23~(1) (1996) 1--10.
\newblock \href {https://doi.org/10.1016/0093-6413(95)00070-4}
  {\path{doi:10.1016/0093-6413(95)00070-4}}.

\bibitem{Wilhelmsen2024}
M.~A. Gjennestad, {\O}.~Wilhelmsen, Thermodynamically consistent modeling of
  gas flow and adsorption in porous media, Int. J. Heat Mass Transfer 226
  (2024) 125462.
\newblock \href {https://doi.org/10.1016/j.ijheatmasstransfer.2024.125462}
  {\path{doi:10.1016/j.ijheatmasstransfer.2024.125462}}.

\bibitem{Kou2026}
J.~Kou, X.~Wang, Energy-stable numerical modeling of coupled gas-water-solid
  processes: {A} semi-implicit discretization of thermodynamically consistent
  governing equations, J. Comput. Phys. 560 (2026) 114925.
\newblock \href {https://doi.org/10.1016/j.jcp.2026.114925}
  {\path{doi:10.1016/j.jcp.2026.114925}}.

\bibitem{Yarushina2026}
V.~Yarushina, Y.~Podladchikov, Thermodynamic foundations of coupled
  thermo-hydro-mechano-chemical processes in geological and geoengineering
  materials with complex rheology, {EGU} General Assembly 2026, Vienna,
  Austria, {EGU}26-3831 (2026).
\newblock \href {https://doi.org/10.5194/egusphere-egu26-3831}
  {\path{doi:10.5194/egusphere-egu26-3831}}.

\bibitem{Liu2026}
L.~Liu, X.~Zhang, X.~Wang, Y.~Qi, Z.~Shi, Elastodynamic equations and wave
  propagation for multiscale wave-induced fluid flow in partially saturated
  media from energy conservation, J. Acoust. Soc. Am. 159~(6) (2026)
  5083--5094.
\newblock \href {https://doi.org/10.1121/10.0044115}
  {\path{doi:10.1121/10.0044115}}.

\bibitem{Zhou2021}
Y.~Zhou, X.~Wang, Y.-y. Dai, Dynamic equation in thermo-piezoelectric
  dissipative media from energy conservation, arXiv preprint (2021).
\newblock \href {http://arxiv.org/abs/2104.12915} {\path{arXiv:2104.12915}}.

\bibitem{Gurtin1981}
M.~E. Gurtin, An Introduction to Continuum Mechanics, Academic Press, New York,
  1981.

\bibitem{Malvern1969}
L.~E. Malvern, Introduction to the Mechanics of a Continuous Medium,
  Prentice-Hall, Englewood Cliffs, NJ, 1969.

\bibitem{Callen1985}
H.~B. Callen, Thermodynamics and an Introduction to Thermostatistics, 2nd
  Edition, John Wiley \& Sons, New York, 1985.

\bibitem{Landau1976}
L.~D. Landau, E.~M. Lifshitz, Mechanics, 3rd Edition, Vol.~1 of Course of
  Theoretical Physics, Pergamon Press, Oxford, 1976.

\bibitem{Cengel2002}
Y.~A. {\c{C}}engel, M.~A. Boles, Thermodynamics: An Engineering Approach, 4th
  Edition, McGraw-Hill, Boston, 2002.

\bibitem{Jackson1999}
J.~D. Jackson, Classical Electrodynamics, 3rd Edition, John Wiley \& Sons, New
  York, 1999.

\bibitem{Landau1960}
L.~D. Landau, E.~M. Lifshitz, Electrodynamics of Continuous Media, Vol.~8 of
  Course of Theoretical Physics, Pergamon Press, Oxford, 1960.

\bibitem{Maugin1988}
G.~A. Maugin, Continuum Mechanics of Electromagnetic Solids, North-Holland,
  Amsterdam, 1988.

\bibitem{Gelfand1963}
I.~M. Gelfand, S.~V. Fomin, Calculus of Variations, Prentice-Hall, Englewood
  Cliffs, NJ, 1963.

\bibitem{Lanczos1970}
C.~Lanczos, The Variational Principles of Mechanics, 4th Edition, University of
  Toronto Press, Toronto, 1970.

\bibitem{Rudin1976}
W.~Rudin, Principles of Mathematical Analysis, 3rd Edition, McGraw-Hill, New
  York, 1976.

\bibitem{Munn1973}
R.~W. Munn, Thermodynamic and physical properties of solids in electric fields,
  J. Phys. C: Solid State Phys. 6~(22) (1973) 3213--3222.
\newblock \href {https://doi.org/10.1088/0022-3719/6/22/008}
  {\path{doi:10.1088/0022-3719/6/22/008}}.

\bibitem{Onsager1931}
L.~{Onsager}, {Reciprocal Relations in Irreversible Processes. I.}, Phys. Rev.
  37~(4) (1931) 405--426.
\newblock \href {https://doi.org/10.1103/PhysRev.37.405}
  {\path{doi:10.1103/PhysRev.37.405}}.

\bibitem{Onsager1931b}
L.~{Onsager}, {Reciprocal Relations in Irreversible Processes. II.}, Phys. Rev.
  38~(12) (1931) 2265--2279.
\newblock \href {https://doi.org/10.1103/PhysRev.38.2265}
  {\path{doi:10.1103/PhysRev.38.2265}}.

\bibitem{Bedford2021}
A.~Bedford, Hamilton's Principle in Continuum Mechanics, Springer, Cham, 2021.
\newblock \href {https://doi.org/10.1007/978-3-030-90306-0}
  {\path{doi:10.1007/978-3-030-90306-0}}.

\end{thebibliography}

\appendix
\crefalias{section}{appendix}
\crefalias{subsection}{subappendix}

\section{Validation via Hamilton's principle}
\label{sec:validation_HP}

To confirm the consistency and thermodynamic robustness of the energy-conservation framework developed in the main context, we now derive the governing equations and boundary conditions using Hamilton's variational principle. This serves as an independent cross-check: since the two methods, i.e., global energy conservation and Hamilton's principle, are fundamentally different in their logical starting points, their yielding of identical field equations and boundary conditions provides strong validation of the proposed theory.

\subsection{Statement of Hamilton's principle}

The variational statement with external loads, i.e.
\begin{equation}
\delta \int_{t_1}^{t_2} \mathcal{L} \, {\rm d}t +
\int_{t_1}^{t_2} {\rm d}t \left[ \int_V f_i^\alpha \delta u_i^\alpha \, {\rm d}V
+ \int_{\partial V} \left( t_i^\alpha \delta u_i^\alpha + \sigma_e \delta \varphi \right) {\rm d}S \right] = 0,
\label{eq:HP_global}
\end{equation}
is the standard generalized Hamilton's principle for non-conservative systems \cite{Lanczos1970, Bedford2021}. In continuum mechanics, the inclusion of body forces and surface tractions is well established \cite{Gurtin1981, Malvern1969}. For piezoelectric media, the electrical terms are treated by \cite{Tiersten1969} and \cite{Maugin1988}, while for multiphase porous media the extension to multiple phases is given by \cite{Coussy2004}.

In \cref{eq:HP_global}, $\mathcal{L} = \int_V \left( \mathcal{T} - \Psi - \rho_f \varphi \right) {\rm d}V$ is the Lagrangian, $\mathcal{T}$ is the kinetic energy density, and $\Psi$ is the material electric enthalpy density. These are exactly the ones derived from the Taylor expansion in \cref{eq:energy_linear}:
\begin{equation}
\mathcal{T} = \frac{1}{2} \rho_{ij}^{\alpha\beta} \dot{u}_i^\alpha \dot{u}_j^\beta,
\qquad
\Psi = \frac{1}{2} C_{ijkl}^{\alpha\beta} \varepsilon_{ij}^\alpha \varepsilon_{kl}^\beta
- e_{ijk}^\alpha E_i \varepsilon_{jk}^\alpha
- \frac{1}{2} \kappa_{ij} E_i E_j,
\label{eq:T_Psi_HP}
\end{equation}
with the quasi-static electric field $E_i = -\varphi_{,i}$. The body force $f_i^\alpha$, the surface traction $t_i^\alpha$ and the surface charge density $\sigma_e$ enter as prescribed external loads, while the prescribed-source coupling $\rho_f \varphi$ is already accounted for in the Lagrangian; consistently, $\rho_f$ is not varied ($\delta\rho_f = 0$).

Taking independent variations of the displacement fields $u_i^\alpha$ and the electric potential $\varphi$, with the variations vanishing at the initial and final times, i.e.
\begin{equation}
\delta u_i^\alpha(\bm{x}, t_1) = \delta u_i^\alpha(\bm{x}, t_2) = 0,
\qquad
\delta \varphi(\bm{x}, t_1) = \delta \varphi(\bm{x}, t_2) = 0,
\label{eq:HP_endpoints}
\end{equation}
and using
\begin{equation}
\delta \dot{u}_i^\alpha = \frac{{\rm d}}{{\rm d}t}(\delta u_i^\alpha),
\qquad
\delta \varepsilon_{ij}^\alpha = \frac{1}{2}(\delta u_{i,j}^\alpha + \delta u_{j,i}^\alpha),
\qquad
\delta E_i = -\delta \varphi_{,i},
\label{eq:HP_variations}
\end{equation}
we integrate by parts in time and space. The variational statement becomes:
\begin{equation}
\begin{aligned}
& \int_{t_1}^{t_2} {\rm d}t \int_V \Big[
\big( -\rho_{ij}^{\alpha\beta} \ddot{u}_j^\beta + \sigma_{ij,j}^\alpha + f_i^\alpha \big) \delta u_i^\alpha
+ \big( D_{i,i} - \rho_f \big) \delta \varphi
\Big] {\rm d}V \\
& + \int_{t_1}^{t_2} {\rm d}t \int_{\partial V} \Big[
\big( t_i^\alpha - \sigma_{ij}^\alpha n_j \big) \delta u_i^\alpha
+ \big( \sigma_e - D_i n_i \big) \delta \varphi
\Big] {\rm d}S = 0,
\end{aligned}
\label{eq:HP_expanded}
\end{equation}

Because the variations $\delta u_i^\alpha$ and $\delta \varphi$ are arbitrary and independent in the volume $V$ and on the boundary $\partial V$, the fundamental lemma of the calculus of variations requires that the coefficients of each variation vanish separately. This yields:

No.1. Volume equations (in the interior $V$):
\begin{equation}
\rho_{ij}^{\alpha\beta} \ddot{u}_j^\beta = \sigma_{ij,j}^\alpha + f_i^\alpha
\qquad \text{(momentum balance)},
\label{eq:HP_momentum}
\end{equation}
\begin{equation}
D_{i,i} = \rho_f
\qquad \text{(Gauss's law)}.
\label{eq:HP_gauss}
\end{equation}
No.2. Natural boundary conditions (on $\partial V$):
\begin{equation}
\sigma_{ij}^\alpha n_j = t_i^\alpha
\qquad \text{(traction condition)},
\label{eq:HP_traction}
\end{equation}
\begin{equation}
D_i n_i = \sigma_e
\qquad \text{(charge condition)}.
\label{eq:HP_charge}
\end{equation}

\subsection{Recovery of the constitutive relations from Hamilton's principle}

We now show explicitly how the constitutive relations of the main text follow from the variational principle once the explicit form of the electric enthalpy density $\Psi$ is known.

The stress tensor $\sigma_{ij}^\alpha$ and the electric displacement $D_i$ were defined as thermodynamic conjugates to the strain and electric field, respectively (see, e.g., refs. \cite{Tiersten1969, Maugin1988}):
\begin{equation}
\sigma_{ij}^\alpha: = \frac{\partial \Psi}{\partial \varepsilon_{ij}^\alpha},
\qquad
D_i: = -\frac{\partial \Psi}{\partial E_i}.
\label{eq:def_sigma_D_HP}
\end{equation}
These definitions are not additional assumptions; they emerge naturally from the boundary terms in the variational principle and are exactly the same as those in the energy-conservation framework (\cref{eq:stress_D_def} of the main text).

The electric enthalpy density $\Psi$ for the linear multiphase piezoelectric medium was derived from the Taylor expansion of the energy density (\cref{eq:energy_linear} of the main text) as
\begin{equation}
\Psi = \frac{1}{2} C_{ijkl}^{\alpha\beta} \varepsilon_{ij}^\alpha \varepsilon_{kl}^\beta
- e_{ijk}^\alpha E_i \varepsilon_{jk}^\alpha
- \frac{1}{2} \kappa_{ij} E_i E_j.
\label{eq:Psi_explicit}
\end{equation}
This same $\Psi$, together with the prescribed-source coupling $\rho_f \varphi$, enters the Lagrangian $\mathcal{L} = \mathcal{T} - \Psi - \rho_f \varphi$ used in Hamilton's principle.

Substituting \cref{eq:Psi_explicit} into the definitions \cref{eq:def_sigma_D_HP} and performing the partial differentiations yields the following.

{For the stress tensor:}
\[
\sigma_{ij}^\alpha = \frac{\partial}{\partial \varepsilon_{ij}^\alpha}
\left[
\frac{1}{2} C_{klmn}^{\beta\gamma} \varepsilon_{kl}^\beta \varepsilon_{mn}^\gamma
- e_{kmn}^\beta E_k \varepsilon_{mn}^\beta
- \frac{1}{2} \kappa_{kl} E_k E_l
\right].
\]
Using the symmetry $C_{ijkl}^{\alpha\beta} = C_{klij}^{\beta\alpha}$ and noting that the third term is independent of strain, we obtain
\begin{equation}
\sigma_{ij}^\alpha = C_{ijkl}^{\alpha\beta} \varepsilon_{kl}^\beta - e_{kij}^\alpha E_k.
\label{eq:sigma_HP}
\end{equation}
{For the electric displacement:}
\[
D_i = -\frac{\partial}{\partial E_i}
\left[
\frac{1}{2} C_{klmn}^{\beta\gamma} \varepsilon_{kl}^\beta \varepsilon_{mn}^\gamma
- e_{kmn}^\beta E_k \varepsilon_{mn}^\beta
- \frac{1}{2} \kappa_{kl} E_k E_l
\right].
\]
The first term does not depend on $E_i$, so
\begin{equation}
D_i = - \left( - e_{imn}^\beta \varepsilon_{mn}^\beta - \kappa_{ik} E_k \right)
= e_{ijk}^\alpha \varepsilon_{jk}^\alpha + \kappa_{ij} E_j.
\label{eq:D_HP}
\end{equation}

Thus, the constitutive relations \cref{eq:sigma_HP,eq:D_HP} are recovered identically -- they are simply the result of differentiating the known potential $\Psi$ with respect to its independent variables. Substituting these relations into the volume equations (momentum balance and Gauss's law) closes the system, exactly as in the energy-conservation method, which confirms that the variational approach yields the same constitutive laws as those derived from the energy-conservation framework.

It is important to distinguish the present approach from earlier works that also employ energy concepts. In classical treatments \cite{Biot1956, Gurtin1981, Malvern1969, Coussy2004}, the use of energy is typically limited to defining kinetic energy, potential energy, or dissipation functions, while Newton's second law is still presupposed as a fundamental postulate. In contrast, the present framework takes the first law of thermodynamics -- i.e., global energy conservation -- as the sole axiomatic starting point. No independent assumption of Newton's second law is made; rather, the momentum equations emerge naturally through the localization of the energy balance. This distinction makes the present approach a genuine first-principles theory, not merely an energy-based reformulation of existing theories, for example, Hamilton's principle.

\end{document}